\documentclass[12pt,letterpaper]{article}

\usepackage{amsmath,amssymb,array,calc,rotating,epsfig,psfrag, amscd}
\usepackage{color}
\usepackage{hyperref}
\usepackage[left=2cm,top=1cm,right=3cm,nohead]{geometry}

\newcommand{\bea}{\begin{eqnarray}}
\newcommand{\eea}{\end{eqnarray}}
\newcommand{\be}{\begin{equation}}
\newcommand{\ee}{\end{equation}}
\newcommand{\barr}{\begin{array}}
\newcommand{\earr}{\end{array}}

\newcommand{\tphi}{\tilde{\phi}}

\newcommand{\txi}{\tilde{\xi}}

\newcommand{\w}{\wedge}

\newcommand{\non}{\nonumber}

\newcommand{\vol}{{\rm vol}}

\newcommand{\bpm}{\begin{pmatrix}}
\newcommand{\epm}{\end{pmatrix}}
 \newcommand{\bitem}{\begin{itemize}}
 \newcommand{\eitem}{\end{itemize}}

\definecolor{cardinal}{rgb}{0.6,0,0}
\definecolor{darkgreen}{rgb}{0,0.5,0}
\definecolor{golden}{rgb}{0.92, 0.7, 0}
\definecolor{midnight}{rgb}{0, 0, 0.5}
\definecolor{darkblue}{rgb}{0.2, 0, 0.8}

\newcommand{\beq}{\begin{equation}\begin{aligned}}
\newcommand{\eeq}{\end{aligned}\end{equation}}
\newcommand{\nn}{\nonumber}

\newcommand{\eee}{{\cal E}}
\newcommand{\fff}{{\cal F}}

\newcommand{\qqq}{{\cal Q}}

\newcommand{\eF}{F(\arcsin(1/r)|-1)}

\begin{document}

\begin{flushright}
IPhT-t11/172
 \end{flushright}

\vspace{0.5cm}
\begin{center}

{\Large \bf Assessing a candidate IIA dual to\\ metastable supersymmetry--breaking}\\
\vskip 3mm

 \vskip1.5cm 
Gregory Giecold, Enrico Goi and Francesco Orsi 
 \vskip0.5cm
$^{*}$\textit{Institut de Physique Th\'eorique,\\
CEA Saclay, CNRS URA 2306,\\
F-91191 Gif-sur-Yvette, France}\\
\vskip0.5cm
gregory.giecold, enrico.goi, francesco.orsi@cea.fr\\
\end{center}
\vskip1.5cm
\begin{abstract}

We analyze the space of linearized non--supersymmetric deformations around a IIA solution found by Cveti\v{c}, Gibbons, L\"{u} and Pope (CGLP) in hep-th/0101096. We impose boundary conditions aimed at singling out among those perturbations those describing the backreaction of anti--D2 branes on the CGLP background. The corresponding supergravity solution is a would--be dual to a metastable supersymmetry--breaking state. However, it turns out that this candidate bulk solution is inevitably riddled with IR divergences of its flux densities and action, whose physical meaning and implications for models of string cosmology call for further investigation.

\end{abstract}

\newpage

\section{Introduction}

Metastable supersymmetry--breaking is an attractive mechanism from a phenomenological point of view~\cite{ISS}. Furthermore, theories for which a metastable supersymmetry breaking state can be realized --- such as $\mathcal{N}=1$, $SU(N_c)$ SQCD in the free magnetic phase with massive flavours --- are relatively simple and generic enough, unlike the comparatively baroque ingredients involved in other approaches to dynamical SUSY--breaking (see for instance~\cite{SS, ISLec} for a review).
 
Attempts have been made to embed the proposal of Intriligator, Seiberg and Shih into string theory (see for instance~\cite{OO,FGEU}), via brane engineering of the electric and magnetic phases~\cite{EGK1}. Nevertheless, in view of the obstruction that seems to arise upon turning on the string coupling $g_s \neq 0$~\cite{BGHSS} or the alternative view~\cite{GK1} that involves string tachyons corrections to argue that the brane configuration still describes the ISS state, it is of interest  to try and find an alternative stringy embedding and search for would--be supergravity duals to metastable supersymmetry--breaking states.

One would (i) start from a well--studied BPS solution of IIB, IIA or eleven--dimensional supergravity, then (ii) add some supersymmetry--breaking ingredients --- typically anti--branes at the bottom of a warped throat --- so as to lift to a de Sitter local minimum of the potential~\cite{KKLT} and obtain by the same token a stringy realization of the inflaton potential~\cite{KKLMMT}. The next step (iii) is to consider the backreaction of such anti--branes on their background. If this can be achieved without any serious singularity or instability, that backreaction procedure would then yield a supergravity dual to a metastable state that is part of the same theory as the vacuum described by the initial, unperturbed supergravity background. 

For instance, a well studied vacuum of type IIB theory is the Klebanov--Strassler solution (KS)~\cite{KS}, which has already less than maximal supersymmetry. It is dual to an $\mathcal{N}=1$ supersymmetric $SU(N+M)\times SU(N)$ gauge theory exhibiting interesting features such as a cascade of Seiberg dualities~\cite{Seiberg:1994pq}, confinement and chiral symmetry breaking.

A realization of metastable SUSY--breaking starting from the KS background has been proposed by Kachru, Pearson and Verlinde (KPV) in the probe approximation~\cite{KPV} (that is, step (ii) from the aforementioned guideline). Supersymmetry is broken by adding a certain amount of anti--branes which are attracted to the bottom of the throat. These authors propose a mechanism in which a fraction of the anti--branes can annihilate (via polarization and the Myers effect~\cite{Myers}) with the positive brane--charge dissolved in flux, a process which is argued to correspond to the decay of the metastable vacuum in the dual field theory description~\cite{KPV}. A related proposal has recently appeared in the work of Klebanov and Pufu~\cite{KP}, in an 11--dimensional supergravity context.

A recent program investigating the construction of metastable states beyond the probe approximation (corresponding to step (iii) above) has been initiated in~\cite{BGrH, BGGH, D, BGGHM1,BGGHM2, BGiH}. The conclusion from that work is that there is an unavoidable singularity in the IR region of the backreacted solutions which have been considered so far.

In order to determine the nature of such singularities, it is worth understanding whether their appearance is related to the particular choice of the background or if it is instead a general feature of the backreaction procedure. There are indeed backgrounds which share enough features with the Klebanov--Strassler background to be candidate setups for arguing about the presence of meta-stable vacua.

In the present paper we will focus our attention on a type IIA setting, for which we will examine the backreaction due to anti--D2 branes. The configuration we will start from is a non--singular fractional+ordinary D2 branes supergravity solution due to Cveti\v{c}, Gibbons, L\"{u} and Pope~\cite{CGLP} (CGLP)\footnote{The supergravity dual to the deconfined phase of the underlying theory has recently been considered by one of us~\cite{Giec}.}.

We find that the candidate IIA supergravity dual to metastable SUSY--breaking\footnote{See~\cite{Giveon:2009bv} for a generalization of the ISS model to lower dimension.} that we build is riddled with singularities arising from the linearized deformation of either the RR or NSNS field strengths. Of much concern, those are non--finite action singularities. A novelty of our work compared to~\cite{BGrH,BGiH} is that those singularities are not sub--leading compared to the kind of singularities that are allowed as a physically sensible ones, that is those stemming from the effect of anti--D2 branes smeared on the $S^4$ at the bottom of the tip.

Whereas for the backreaction of anti--D3's on the Klebanov--Strassler solution one could have expected, with hindsight, a singularity to arise in analogy with the IIA brane engineering of four--dimensional gauge theories, a similar argument does not hold for string theory constructions of 2+1--dimensional gauge theories.

Indeed, the profile of the NS5--branes featured in those brane engineering constructions is generally not rigid but is instead sourced by the stack of Dp branes in--between (see~\cite{Giveon:1998sr} for pointers to the literature and much more on the physics of those brane constructions). For four--dimensional field theories living on D4--branes between two NS5's, the profile determined upon solving a Laplace equation is logarithmically running. This corresponds to the log--running of the gauge coupling for asymptotically free theories. 

On the other hand, for three--dimensional field theories living on D3--branes between two NS5's, the profile decays as $1/r$ away from the location of the D3's on the NS5. Such a mode does not have the potential ability to enhance small IR fluctuations into log--running ones, an ability to which one might roughly ascribe the singularities encountered in the holographic approach to realizing metastable states in string theory, if those singularities are deemed as truly pathological\footnote{We keep all the options open on this issue. Another comment on this appears in section~\ref{SAD2}.}. 

So, proceeding in analogy with brane engineering constructions, for 2+1--dimensional IR perturbations should be expected not to affect the UV asymptotics of the background. As we shall see as an outcome of our linearized deformation analysis, this is not quite the case for the candidate supergravity dual to a 2+1--dimensional metastable state. The IR singularities we find are affecting the UV behavior, in the sense that they cannot be completely tamed without switching off at the same time the force felt by a probe D2--brane in the UV.

Besides, having their legs in the wrong directions, those IR divergences cannot be identified as the remnant signature of an NS5 instanton through which the metastable state is been argued to decay in the probe approximation~\cite{KPV, KP}. 

Such singularities cannot be identified either with those characterizing fractional branes on Ricci--flat transverse geometries before the resolution or deformation of those manifolds (solutions of the Klebanov--Tseytlin~\cite{KT} type, whose singularities get resolved in the Klebanov--Strassler solution).

The situation is quite puzzling and it might well be that those singularities are an artifact of having to smear anti--branes in order to make the problem tractable. Some recent results~\cite{Bla1, Bla2} suggest however that a localization procedure is bound to make things worse, rather than alleviating them. Furthermore, an analysis at full non--linear level~\cite{Bla3} gives evidence of how a divergent behaviour is still present even when completely localized sources are considered. 

It is therefore a reasonable alternative viewpoint consider that such Coulomb--like divergencies as physically meaningful, and might somehow be used to discriminate among solutions of the string theory landscape.

We propose the following analogy for linking an eventual singular behaviour and its physical causes. In QCD, there are free quarks in the linearized approximation. Their ``backreaction'' results in a Coulomb--like singularity. We know that this is an indication that quarks are not good approximations at all to finite--energy states from the spectrum of QCD, which instead consists of confined, colorless states.

Now, the singularities we find involve in particular an IR--divergent NSNS flux--density. We would like to suggest, following the above--mentioned situation in QCD, that those singularities perhaps hint that some of the scenarios that have been proposed in a probe approximation to uplift an AdS--vacuum to a metastable de Sitter one using brane sources do not engineer acceptable states of the ``spectrum'' of string theory. But how is the analogy to hold with the Coulomb--like singularities of QCD, given that (anti--) branes are not expected to source NSNS flux ? Or do they ?

Indeed, they certainly do not if we stick to the guideline and intuition drawn from the supergravity solutions describing such sources in flat space. Nevertheless, it has been proposed that they naturally do~\cite{D,Shiu} in perturbation theory around a complicated, warped geometry such as the Klebanov--Strassler solution or the CGLP solution we investigate in the present paper. 

Besides, the authors of~\cite{D,Shiu} argue rather convincingly that the IR--divergencies that seem to affect, at linearized order in the SUSY--breaking parameter, the backreaction by antibranes of an underlying warped background should disappear at full non--linear order. 
The claim goes as follows: the singularities in the flux densities are naturally sourced at linear--order of perturbation theory by the acceptable $1/\tau$ behavior of the deformation of the warp factor due to, say, the bunch of anti--D3's smeared on the $S^3$ of the Klebanov--Strassler geometry. It is then advocated that the $1/\tau$ contribution of first--order perturbation theory\footnote{$\tau$ denotes the radial variable in the bulk.}, summed up with the $1/\tau^2$ contribution at second--order, and so on with all the other contributions, are nothing but terms in the expansion of the inverse warp factor of the backreacted background modified by the presence of antibranes. The whole sum of the individual contributions at each order is then claimed to be a perfectly regular quantity, $h^{-1}$, the inverse of the backreacted warp factor. As a result, it is argued that singularities in the fluxes are an artifact of perturbation theory that should wash out at full non--linear order. 

It is currently a daunting task to check if this possibility is indeed realized. However, we link on this possibility in section~\ref{SAD2}, where we argue how the issue of those singularities in the smearing approximation could maybe be settled by considering 2nd--order expansions for the deformation modes of a BPS background, a task which has not been attempted so far.

But, to come back to the tentative analogy with QCD, it might seem after, in view of the above chain of argument involving the full backreaction by antibranes (something we do not attempt; we stick at linearized deformation), that this analogy does not hold when applied to the backreaction of antibranes at full non--linear order. So what ?

The reason we maintain that this analogy might possibly hold is that the afore--mentioned argument seeking to explain how the singularities in the fluxes should vanish at full non--linear order is not entirely water--tight. Indeed, a recent paper~\cite{Bla3} very convincingly shows that in some instance a singularity in the $H_3$ flux density is still present at full non--linear order !  

It is beyond the scope of the present work to offer more credence to vindicate or dispel the following possibility but it is very tempting to imagine that the IR singularities we keep on finding upon backreacting the effect of antibranes on some BPS background are similarly a hint that some of the constructions which have been proposed has duals to metastable SUSY--breaking might instead belong to some ``swampland''~\cite{swamp} once the backreaction of the SUSY--breaking ingredients is duly taken into account.\\

Our results are organized as follows. In section~\ref{SCGLP}, we review the CGLP solution, linearized perturbations of which pervade the bulk of the present work. We then recall the basics of the method developed by Borokhov and Gubser in section~\ref{SBG}. As our analysis makes extensive use of a superpotential for the CGLP background, we also outline how the latter is derived, as first found by Herzog~\cite{H}. Next, we explicitly evaluate the force acting on a probe D2--brane in section~\ref{SF}. For a lighter reading, a key part of our work has been relegated to appendix~\ref{Sspace}, where we expose solutions to one of the system of coupled ODE's governing the supersymmetry--breaking modes. As for the remaining modes, their IR asymptotics appear in appendix~\ref{Sspace} as well. In order to calibrate all that machinery, section~\ref{SBPS} pertains to adding BPS D2 branes to the CGLP background. Finally, in section~\ref{SAD2} we impose IR boundary conditions appropriate to the backreaction of anti--D2 branes smeared at the tip of that warped throat background. We explain how an IR--divergences in the RR or NSNS flux inevitably comes about.

\section{Ansatz for the perturbation}\label{SCGLP}

We start with explaining the Ansatz under consideration in this paper. It respects the symmetries of the supersymmetric regular + fractional D2 branes supergravity solution found by Cveti\v{c}, Gibbons, L\"{u} and Pope. Non--supersymmetric, linearized deformations of the CGLP background are part of this Ansatz. 

In Einstein frame, it is given by
\beq
& ds_{10}^2 = e^{-5 z(r)}\, \eta_{\mu \nu}\, dx^{\mu}\, dx^{\nu} + \ell^2\, e^{3 z(r)}\, \left[ h(r)^2\, dr^2 + e^{2 u}\, \left( D\mu^i \right)^2 + e^{2 v}\, d\Omega_4 \right]\, ,\nonumber
\eeq
\begin{align}\label{Ansatz}
g_s\, F_4 =&\, K(r)\, d^3 x \wedge dr + 2\, m\, \left( g_1(r) + c_2 \right)\, J_2 \wedge J_2 + 2\, m\, \left( g_1(r) + c_3 \right)\, U_2 \wedge J_2 \nonumber\\ & + m\, g_1^{\prime}(r)\, \epsilon_{ijk} \mu^i\, dr \wedge D\mu^j\wedge J^{k} \, ,
\end{align}
\beq
\ell \, B_2 = m\, \left[ g_2(r)\, U_2 + g_3(r)\, J_2\right] \, , \qquad F_2 = 0 \, , \qquad \Phi = \Phi(r) \, ,\nonumber
\eeq
where $d\Omega_4 = e^{\alpha} e^{\alpha}$ and 
\beq
D\mu^i = d\mu^i + \epsilon_{ijk}\, A^j\, \mu^k
\eeq
denote line elements on the $\text{S}^2$ fiber over the $\text{S}^4$ base. The coordinates $\mu^i$, $i = 1,2,3$ obey the constraint $\mu^i\, \mu^i = 1$. The $A^j = A^j_{\alpha}\, e^{\alpha}$ refer to $su(2)$ Yang--Mills instanton potentials. Their field strength components
\beq
J^i = dA^i + \frac{1}{2}\, \epsilon^{i}_{\, jk}\, A^j \wedge A^k
\eeq
satisfy the algebra of the unit quaternions, i.e. 
\beq\label{quaternions}
J^i_{\alpha \gamma}\, J^j_{\gamma \beta} = - \delta^{ij}\, \delta_{\alpha \beta} + \epsilon^{ij}_{\, \, k}\,  J^k_{\alpha \beta} \, .
\eeq
All in all, this makes the seven--dimensional transverse space a cone over a squashed $\mathbb{CP}^3$~\cite{GPP}. 

The Bianchi identities are satisfied in view of the following definitions and identities:
\begin{align}
& U_2 \equiv \frac{1}{2}\, \epsilon_{ijk}\, \mu^i\, D\mu^j \wedge D\mu^k \, , \qquad J_2 \equiv \mu^i\, J^i \, , \qquad U_3 \equiv D\mu^i \wedge J^i \, ,\nonumber\\&
\, \, \, \, \, \, \, \, \, \, \, \, \, \, \, \, \, \, \, \, \, \, \, \, \, \, \, \, \, \, \, \, dU_2 = U_3 \, , \qquad d J_2 = U_3 \, , \qquad d U_3 = 0 \, .
\end{align}

\subsection{The underlying superpotential}

We are next going to reduce to a one--dimensional sigma model the bosonic part of the $\text{IIA}$ supergravity action. The reason for doing so is as follows: the machinery we rely on in order to conveniently obtain the linearized non--supersymmetric deformations around a supersymmetric background involves a superpotential. By a superpotential, here we mean a convenient book--keeping scalar combination of the fields entering a given supergravity Ansatz. By definition, field--space derivatives of the superpotential times the sigma--model metric yield the potential that is obtained upon reducing the higher--dimensional supergravity Ansatz to the one--dimensional sigma model.

In Einstein frame, the IIA action reads 
\begin{align}\label{SUGRAIIaEinsteinFrame}
\mathcal{S}_{IIA} =&\, \frac{1}{2\, \kappa^2} \int d^{10}\, x\, \sqrt{\mid g \mid}\,  R - \frac{1}{4\, \kappa^2} \int \Big[ d\Phi \wedge \star d\Phi + g_s\, e^{-\Phi}\, H_3 \wedge \star H_3 \nonumber\\ & + g_s^{1/2}\, e^{3 \Phi/2}\, F_2 \wedge \star F_2 + g_s^{3/2}\, e^{\Phi/2}\, \tilde{F}_4 \wedge \star \tilde{F}_4 + g_s^2\, B_2 \wedge F_4 \wedge F_4 \Big]\, ,
\end{align}
where 
\beq
\tilde{F}_4 = F_4 - C_1 \wedge H_3 \, , \qquad F_4 = d C_3 \, , \qquad H_3 = d B_2\, , \qquad F_2 = d C_1 \, .
\eeq
Inserting the above expressions for the fields and metrics~\eqref{Ansatz} yields
\beq
\mathcal{S}_{\text{IIA}} = \frac{\ell^5 \, \text{Vol}\left(M_{1,2}\right)\, \text{Vol}\left( M_6 \right)}{2\, \kappa^2}\, \int dr\, \mathcal{L} \, , 
\eeq
where $\mathcal{L} = T - V$ and $\text{M}_{1,2}$, $\text{M}_{6}$ denote the 2+1 dimensional Minkowski space and the level surfaces of the seven--dimensional $\text{G}_2$--holonomy manifold, respectively. 

The kinetic term is
\begin{align}\label{kinetic}
T =&\, \frac{e^{2 u + 4 v}}{h}\, \Big[ -30\, z^{\prime \, 2} + 2\, u^{\prime \, 2} + 12\, v^{\prime \, 2} + 16\, u^{\prime}\, v^{\prime} - 2\, g_s^{-1/2}\, \frac{m^2}{\ell^6}\, e^{-9 z + \Phi/2 - 2 u - 4 v}\, g_1^{\prime \, 2} \nonumber\\ & - \frac{g_s}{2}\, \frac{m^2}{\ell^6}\, e^{-6 z - \Phi}\, \left( g_2^{\prime \, 2}\, e^{-4 u} + 2\, g_3^{\prime \, 2}\, e^{-4 v} \right) - \frac{1}{2} \Phi^{\prime \, 2}  \Big]
\end{align}
and, as anticipated, it is clear that both $h(r)$ and $K(r)$ are non--dynamical fields. 

Then, upon eliminating the non--dynamical $K$ through its algebraic equation of motion, namely
\beq
K=\frac{4m^2}{\ell^6}g_s^{1/2}e^{-2u-4v-15z-\Phi/2}\,h\,\Big[g_1(g_2+g_3)+c_2g_2+c_3g_3\Big]\label{Kvalue}
\eeq
and evaluating the Lagrangian at the corresponding minimum for $K$, the potential becomes
\begin{align}\label{potential}
V =&\, - 2\, h\, e^{- 2 u - 4 v}\, \left[ e^{2 u + 8 v} - e^{6 u + 4 v} + 6\, e^{4 u + 6 v} \right] + 2\, g_s\, h\, \frac{m^2}{\ell^6}\, e^{-6 z - \Phi} \left[ g_2 + g_3 \right]^2 \nonumber\\ & + 4\, g_s^{-1/2}\, \frac{m^2}{\ell^6}\, e^{- 9 z + \Phi/2 + 2 u}\, h\, \left[ 2\, \left( g_1 + c_2\right)^2\, e^{-4 v} + \left( g_1 + c_3 \right)^2\, e^{-4 u} \right] \nonumber\\ & + 8\, g_s^{1/2}\, \frac{m^4}{\ell^{12}}\, e^{-15 z - \Phi/2 - 2 u - 4 v}\, h\, \left[ g_1\, \left( g_2 + g_3 \right) + g_2\, c_2 + g_3\, c_3 \right]^2 \, .
\end{align}

Writing the Lagrangian as 
\beq
\mathcal{L} = - \frac{1}{2}\, G_{ab}\, (d{\phi^a}/d r)\, 
(d{\phi^b}/d r) - V \, ,
\eeq
where we denote the set of functions $\phi^{a}$, $a = 1,\, ...,\, 7$ in the following order
\beq\label{phi a set}
\phi^{a} = \left( u, v, z, \Phi, g_1, g_2, g_3 \right) \, ,
\eeq 
we find that the following superpotential, initially found by Herzog~\cite{H},
\beq\label{spptl}
W = - 8\, \left[ e^{u + 4 v} + e^{3 u + 2 v} \right] + 8\, \frac{m^2}{\ell^6}\, g_s^{1/4}\, e^{-\frac{15}{2} z -\frac{\Phi }{4}}\, \left[ g_1\, \left( g_2 + g_3 \right) + g_2\, c_2 + g_3\, c_3 \right] \nonumber\\
\eeq 
correctly accounts for all the terms in the potential~\eqref{potential}, that is to say
\beq\label{Vgr}
V = \frac{1}{8} \, G^{ab} \, \frac{\partial W}{\partial \phi^a}\, \frac{\partial W}{\partial \phi^b} \, .
\eeq
One can check that the zeroth--order CGLP solutions that we are about to summarize below obey the first--order BPS equations derived from this superpotential:
\beq
\frac{d\phi^a}{dr} - \frac{1}{2}\, G^{ab}\, \frac{\partial W}{\partial \phi^b} = 0 \, .
\eeq
This motivates the respective choice of signs in front of the metric part and the flux pieces of the superpotential~\eqref{spptl}, which are otherwise arbitrary.

\subsection{Zeroth--order solution}

The solution of Cveti\v{c}, Gibbons, L\"{u} and Pope corresponding to resolved fractional D2--brane with transverse seven--dimensional squashed cone over $\mathbb{CP}^3$ preserves $1/4$ of the original supersymmetry, giving rise to a dual $\mathcal{N} = 1$ field theory in 2+1 dimensions.

Let us now gather the expressions for the zeroth--order functions entering this solution, around which we will next expand. It might be appropriate to remind that the radial variable $r$ runs from one to infinity\footnote{\label{fconv}From now on, superscripts will refer to the perturbation order, while subscripts label different functions; quantities which are not labelled by a superscript do not enter the set of perturbed scalars which we introduced in~\eqref{phi a set}.}.
\begin{align}\label{0th solt}
& h= \left( 1 - \frac{1}{r^4} \right)^{-1/2} \, , \qquad e^{2\, u^{0}} = \frac{1}{4}\, r^2 \left( 1 - \frac{1}{r^4} \right)\, , \qquad e^{2\, v^{0}} = \frac{1}{2}\, r^2 \, , \nonumber\\
&g_1^{0} =  \int_1^{r} f_1(y) \, dy \, , \qquad f_1 = e^{u^{0} + 2\, v^{0}}\, u_1 \, , \qquad u_1 = \frac{1}{4\, r^4\, \left( r^4 - 1 \right)} - \frac{\left( 3\, r^4 - 1 \right)\, P(r)}{4\, r^5\, \left( r^4 - 1 \right)^{3/2}} \, , \nonumber\\
&g_2^{0}= \int_1^{r} f_2(y)\, dy \, , \qquad f_2 = h\, e^{2\, u^{0}}\, u_2 \, , \qquad u_2 = \frac{1}{r^4} + \frac{P(r)}{r^5\, \left(r^4 -1\right)^{1/2}} \, , \nonumber\\
&g_3^{0} =  \int_1^{r} f_3(y) \, dy \, , \qquad f_3 = h\, e^{2\, v^{0}}\, u_3 \, , \qquad u_3 = - \frac{1}{2\, \left( r^4 - 1 \right)} + \frac{P(r)}{r\, \left( r^4 - 1 \right)^{3/2}} \, ,
\end{align}
along with
\begin{align}\label{0th z}
& H_0 \equiv e^{8\, z^0} = \frac{m^2}{\ell^6} \int_r^{\infty} y^5\, \left[ u_3(y) - u_2(y) \right]\, u_1(y) \, dy \, ,\\
& e^{\Phi^{0}} = g_s H_0^{1/4} \, ,
\end{align}
where 
\begin{align}
P(r) &= \int_1^{r} \frac{d u}{\sqrt{u^4 - 1}} \, , \nonumber\\
& = K(-1) - F(\arcsin(1/r) \mid -1) \, .
\end{align}
From now on, $F(\phi \mid k)$ denotes an incomplete elliptic integral of the first kind and $K(k) = F(\pi/2 \mid k)$. We also will encounter elliptic integrals of the second kind $E(\phi \mid k)$. See Appendix~\ref{ellfunc} for a quick reminder. 

The expression for the warp factor $H_0$ above\footnote{It is straightforward to check that the expression~\eqref{0th z} for the warp factor is identical to the one provided by Herzog in~\cite{H}, i.e.~$H_0 = \frac{m^2}{2\, \ell^6}\, \int_r^{\infty} y\, \left[ 2\, u_3(y) - 3 \right]\, u_1(y)\, dy \, ,$ taking into account different conventions.} arises from the condition on the trace of Einstein's equations
\begin{align}\label{box H}
\Box H_0 &= - \frac{1}{6}\, m^2 \mid G_3^{0} \mid^2 \, , \nonumber\\
& = - \frac{1}{6}\, m^2 \mid G_4^{0} \mid^2 \, ,
\end{align}
where we generally define $G_4$ via $F_4 \equiv K \, d^3x \wedge dr + m\, G_4$ and we have used the fact that $G_4^{0} = \star_7\, G_3^{0}$ for the zeroth--order solution.
This can be integrated to
\beq
\star_{10} \left( e^{\Phi^0 / 2}\, d^3 x \wedge dH_0^{-1} \right) = - g_s^{-1/2}\, m\, G_4^0 \wedge B_2^0 \mid_{M_6} \, , 
\eeq
with $M_6$ the level surface of the $G_2$ holonomy manifold which is part of the CGLP solution and ensures its regularity.

Note that the UV behavior of the warp factor is as expected, namely 
\beq
H_0(r) = \frac{Q}{r^5} + {\cal O}(1/r^6) \, , \qquad r \rightarrow \infty \, .
\eeq 
The IR asymptotics of $H$ is
\beq
H_0(r) = H_0(1) - \frac{7}{16}\, \frac{m^2}{\ell^6}\, (r-1) + {\cal O}(r-1)^2 \, , \qquad r \rightarrow 1 \, ,
\eeq
which establishes that $r = 1$ is a coordinate singularity and that the metric is actually regular there. Indeed, notice that in the IR the unperturbed ten--dimensional metric takes the form
\beq
ds_{10}^2=H_0(1)^{-5/8}ds_{\mbox{\tiny{$Mink_3$}}}^2+H_0(1)^{3/8}\left[\frac{1}{4(r-1)}dr^2+(r-1)(D\mu^i)^2+\frac{1}{2}d\Omega_4^2  \right] \, .
\eeq
The coordinate singularity at $r=1$ can be eliminated by shifting gears to 
\beq\label{tau to r}
\tau \equiv  \sqrt{r-1} \, .
\eeq 
The space transverse to the D2 branes therefore approaches $\mathbb{R}^3 \times S^4$ in the far IR. 

As for the constants $c_2$ and $c_3$ appearing in our Ansatz~\eqref{Ansatz}, the background only specifies their difference
\beq\label{c2c3}
c_2 - c_3 = \frac{3}{32} \, .
\eeq

Amazingly, it turns out that $g_1^0$, $g_2^0$ and $g_3^0$ can written in terms of the functions $f_1$ and $f_2$ or $f_3$ appearing in~\eqref{0th solt}: 
\beq\label{g23 simplified}
g_2^0 = - 8\, e^{2\, u^0 + 2\, v^0}\, f_1\, , \qquad g_3^0 = \left( 1 + 8\, e^{2\, u^0 + 2\, v^0} \right)\, f_1 \, .
\eeq
\beq
g_1^0 = \frac{1}{4\, h^0}\, e^{-2\, u^0 + 4\, v^0}\, f_2 - c_2 \, \, \, \, \, \, \text{or equivalently} \, \, \, \, \, \, g_1^0 = \frac{1}{2\, h^0} e^{2\, u^0}\, f_3 - c_3 \, . 
\eeq
Note that those modes are well--behaved in the IR and their series expansions go as
\bea
&& g_1^0 + c_2 = \frac{3}{32} - \frac{1}{16}\, \left( r - 1 \right) + {\cal O}(r-1)^2 \, , \qquad g_2^0 = \frac{1}{2}\, (r-1)^{3/2} + {\cal O}(r-1)^{5/2} \, , \non\\ && g_3^0 = - \frac{1}{8}\, (r-1)^{1/2} - \frac{3}{160}\, (r-1)^{3/2} + {\cal O}(r-1)^{5/2} \, ,
\eea
which might not be immediately obvious from their defining formulae. Actually, the radial derivatives of $g_3$ and $g_2$ also make their way in $H_3$. Correspondingly, a piece of the NSNS flux behaves as $\frac{1}{\sqrt{r-1}}\, dr\w J_2$ in the infrared. This does not signal a pathological behavior and instead is just another instance of the unphysical and non--pathological, coordinate singularity $h^{0}\, dr^2 \sim \frac{dr^2}{r-1}$ that we have already encountered.

\section{The Borokhov--Gubser method}\label{SBG}

The method proposed by Borokhov and Gubser in~\cite{BG} allows to find non--supersymmetric supergravity solutions, starting from a given BPS background. The idea behind the technique is as follows: rather than having to solve $n$ second--order equations for the $n$ fields $\phi_a$ entering a supergravity Ansatz encompassing the background solution, we trade those complicated 2nd--order differential equations for $2\, n$ first--order equations governing those fields $\phi_a$ and their  ``canonical conjugate variables'' $\xi^a$. 

The simplicity of the method has much to do with the fact that $\xi^a$ always form a close system. The equations for the modes of much physical interest, $\phi_a$ involve the conjugate modes $\xi^a$ as source terms. Let us quickly review that approach of Borokhov and Gubser.

We rewrite the Lagrangian by means of the superpotential~\eqref{spptl} as follows
\beq\label{Lagrangian_phi}
 \mathcal{L} =-\frac{1}{2}\, G^{ab}\, \left( \frac{d\phi_a}{d r} - \frac{1}{2}\, G_{ac}\, \frac{\partial W}{\partial\phi_c} \right)\, \left( \frac{d\phi_b}{d r} - \frac{1}{2}\, G_{bd}\, \frac{\partial W}{\partial\phi_d} \right) - \frac{1}{2}\, \frac{d W}{d r} \, .
\eeq
The gradient flow equations obeyed by the underlying BPS solution\footnote{In the case of present interest this BPS solution is the CGLP solution that we have introduced in Section 2.2.} read
\beq\label{gradflow}
 \frac{d\phi_a}{d r}=\frac{1}{2}G_{ab}\frac{\partial W}{\partial \phi_b}\, .
\eeq
Furthermore, the ``zero-energy'' condition arising from the $G_{rr}$ Einstein equation is a constraint that applies to any solution, BPS or not:
\beq\label{zero-energy}
-\frac{1}{2}G_{ab}\frac{d\phi_a}{d r}\frac{d\phi_b}{d r} + V(\phi)= 0 \, .
\eeq

The method of Borokhov and Gubser~\cite{BG} relies on a superpotential to determine perturbations to a solution of~\eqref{gradflow} that satisfy the equations of motion but not necessarily~\eqref{gradflow} itself.  Let us consider an expansion of the fields $\phi_a$ around their supersymmetric value $\phi_a^{0}$,
\beq\label{phiexpansion}
 \phi_a=\phi_a^{0}+\phi_a^{1}(\alpha)+{\cal O}(\alpha^2)
\eeq
for some set of parameters $\alpha$. Let us introduce the following notation
\beq
\xi^a=G^{ab}(\phi^{0})\left(\frac{d \phi_b^{1}}{d r}-N_b^{\phantom{5}d}(\phi_0)\phi_d^{1}\right) \, , \qquad \mathrm{where}\qquad N_b^{\phantom{5}a}(\phi^{0})=\frac{1}{2}\frac{\partial}{\partial\phi_a}\left( G_{bc}\frac{\partial W}{\partial\phi_c}\right)\,.
\eeq
Inserting the expansion~\eqref{phiexpansion} into the equations of motion derived from the one--dimensional Lagrangian, and keeping terms up to the linear order, one obtains
\begin{align}
	\frac{d\xi^a}{d r}+\xi^b\,N_b^{\phantom{5}a}(\phi^{0})&=0\,,\label{xi_eqs}\\
	\frac{d\phi_a^{1}}{d r}-N_a^{\phantom{5}b}(\phi^{0})\phi_b^{1}&=G_{ab}(\phi^{0})\xi^b\label{phi_eqs}\,,
\end{align}
while the constraint~\eqref{zero-energy} can be written as
\beq\label{zero_energy_perturb}
\xi^a\frac{d\phi_a^{0}}{d r}=0\,. 
\eeq
The functions $\xi^a$ are a measure of the deviation from the gradient flow equations~\eqref{gradflow}. Notice that for a supersymmetric deformation all the $\xi^a$ vanish. The obvious advantage of this method is that one can solve separately for the first--order subsystem~\eqref{xi_eqs} and then solve for~\eqref{phi_eqs} which are again first--order.

\subsection{The first--order equations for the supersymmetry--breaking deformations}

\subsubsection{$\tilde{\xi}$ equations}\label{Sxi}

We present the system~\eqref{xi_eqs} of first--order equations for the fields $\xi^a$, which are conjugate to the linearized deformations $\phi_a$ of the CGLP background, in terms of a convenient change of variables
 \beq
\tilde{\xi}_a = \left(\xi_1, \, \xi_1 - \xi_2, \, \xi_3 + 2\, \xi_4, \, \xi_4, \, \xi_5, \, \xi_6, \, -\xi_6 + \xi_7 \right)\, .\label{xitrans}
\eeq
The above combinations were chosen so as to make the corresponding system of equations much easier to solve. We actually managed to find fully analytic expressions for the $\xi^a$ conjugate modes.

The equations are listed in the order in which we have solved them:
\begin{align}
\txi_3'&=-4\frac{m^2 g_s^{1/4}}{l^6}  h\, e^{-2u^0-4v^0-\frac{15z^0}{2}-\frac{\Phi^0}{4}} \Big[c_2
   g^0_2+c_3g^0_3+g^0_1 (g^0_2+g^0_3)\Big]\txi_3 \label{xit3}\\ 
   \nonumber\\
\txi_7'&=-\frac{3m^2g_s^{1/4}}{64 l^6} h\,  e^{-2 u^0-4 v^0-\frac{15 z^0}{2}-\frac{\Phi^0}{4}}  \txi_3\\
   \nonumber\\
\txi_5'&=-\frac{1}{2g_s^{3/4}l^6} h\, e^{-2u^0-4v^0-\frac{15z^0}{2}-\frac{\Phi^0}{4}} \Big[4l^6e^{4v^0+6z^0+\Phi^0}(\txi_6+\txi_7)+8l^6e^{4u^0+6z^0+\Phi^0}\txi_6\nonumber\\
          &\,\,\,\,\,\,\,-g_s m^2 (g_2^0+g_3^0)\txi_3\Big]\\
    \nonumber\\
\txi_6'&=\frac{g_s^{1/4}}{2l^6}h\,e^{-2u^0-4v^0-\frac{3}{4}(10z^0+\Phi^0)}\Big[-2g_s^{1/2}l^6e^{2u^0+4v^0+9z^0} \txi_5+e^{\frac{\Phi^0}{2}}m^2(c_2+g^0_1)\txi_3\Big]\\
    \nonumber\\
\txi_4'&=\frac{h}{8g_s^{3/4}}e^{-\frac{3}{4}(10z^0+\Phi^0)}\Big[-24 e^{2u^0-4v^0+6z^0+\frac{3}{2}\Phi^0}(c_2+g^0_1)\txi_6\nonumber\\
	 &\,\,\,\,\,\,\,-12e^{-2u^0+6z^0+\frac{3}{2}\Phi^0}(c_3+g^0_1)(\txi_6+\txi_7)+6e^{9z^0}g_s^{3/2}(g_2^0+g_3^0)\txi_5\nonumber\\
	 &\,\,\,\,\,\,\,-\frac{m^2g_s}{l^6}e^{-2u^0-4v^0+\frac{\Phi^0}{2}}(c_2g_2^0+c_3g_3^0+g^0_1(g^0_2+g^0_3))\txi_3\Big]
   \end{align}
   \begin{align}
\txi_1'&=\frac{1}{g_s^{3/4}l^6}h\,e^{-2u^0-4v^0-\frac{15}{2}z^0-\frac{\Phi^0}{4}}\Big[g_s^{3/4}l^6e^{u^0+4v^0+\frac{15}{2}z^0+\frac{\Phi^0}{4}}\txi_1+g_s^{3/4}l^6e^{\frac{1}{4}(12u^0+8v^0+30z^0+\Phi^0)}\txi_2\nonumber\\
	 &\,\,\,\,\,\,\,-8l^6e^{4u^0+6z^0+\Phi^0}(c_2+g_1^0)\txi_6+4l^6e^{4v^0+6z^0+\Phi^0}(c_3+g_1^0)(\txi_6+\txi_7)\nonumber\\
	 &\,\,\,\,\,\,\,-g_sm^2(c_2g_2^0+c_3g_3^0+g^0_1(g^0_2+g^0_3))\txi_3\Big]\end{align}
	 \begin{align}\txi_2'&=\frac{1}{g_s^{3/4}l^6}h\,e^{-2u^0-4v^0-\frac{15}{2}z^0-\frac{\Phi^0}{4}}\Big[g_s^{3/4}l^6 e^{u^0+4v^0+\frac{15}{2}z^0+\frac{\Phi^0}{4}}\txi_1+3g_s^{3/4}l^6e^{\frac{1}{4}(12u^0+8v^0+30z^0+\Phi^0)}\txi_2\nonumber\\
	 &\,\,\,\,\,\,\,-24l^6e^{4u^0+6z^0+\Phi^0}(c_2+g_1^0)\txi_6+4l^6e^{4v^0+6z^0+\phi^0}(c_3+g_1^0)(\txi_6+\txi_7)\nonumber\\
	 &\,\,\,\,\,\,\,+g_sm^2(c_2g_2^0+c_3g_3^0+g^0_1(g^0_2+g^0_3))\txi_3\Big]\label{xit2}
\end{align}

\subsubsection{$\tilde{\phi}$ equations}\label{Sphi}
 As previously done for the $\tilde{\xi}$ equations, we shift the original $\phi$ to a more tractable linear combination $\tphi$, defined as\footnote{The inverse transformation is 
\begin{align}
\phi^{a} =&\, \Big( \tilde{\phi}_1, \, \frac{1}{2}\, \left( \tilde{\phi}_1 - \tilde{\phi}_2 \right), \, -\frac{7}{12}\, \tilde{\phi}_1 + \frac{1}{4}\, \tilde{\phi}_2 + \frac{1}{96}\, \tilde{\phi}_3 + \frac{1}{32}\, \tilde{\phi}_4, \, \frac{3}{2}\, \tilde{\phi}_1 + \frac{1}{2}\, \tilde{\phi}_2 - \frac{5}{16}\, \tilde{\phi}_3 + \frac{1}{16}\, \tilde{\phi}_4, \nonumber\\ & \, \, \, \, \, \, \tilde{\phi}_5, \, \frac{1}{2}\, \left( \tilde{\phi_6} + \tilde{\phi}_7 \right), \, \frac{1}{2}\, \left( \tilde{\phi_6} - \tilde{\phi}_7 \right) \Big)\,. \label{inversephi}
\end{align}}
\beq
\tilde{\phi}_{a} = \left( \phi_1, \, \phi_1 - 2\, \phi_2, \, 8\, \phi_1 + 6\, \phi_3 - 3\, \phi_4, \, 8\, \phi_1 + 16\, \phi_2 + 30\, \phi_3 + \phi_4, \, \phi_5, \, \phi_6 + \phi_7, \, \phi_6 - \phi_7 \right) \, .\label{phitrans}
\eeq

The set of equations \eqref{phi_eqs} explicitly reads
\beq\label{phi1}
\tilde{\phi}_{1}^{\prime} = \frac{1}{20}\, h\, e^{-2 u^0 - 4 v^0}\, \left[ \tilde{\xi}_1 + 2\, \tilde{\xi}_2 - 20\, e^{u^0 + 4 v^0}\, \tilde{\phi}_1 - 20\, e^{3 u^0 + 2 v^0}\, \tilde{\phi}_2 \right] \, ,
\eeq

\beq\label{phi2}
\tilde{\phi}_{2}^{\prime} = \frac{1}{20}\, h\, e^{-2 u^0 - 4 v^0}\, \left[ 4\, \tilde{\xi}_1 + 3\, \tilde{\xi}_2 - 20\, e^{u^0 + 4 v^0}\, \tilde{\phi}_1 - 60\, e^{3 u^0 + 2 v^0}\, \tilde{\phi}_2 \right] \, ,
\eeq

\begin{align}\label{phi3}
\tilde{\phi}_{3}^{\prime} =&\, \frac{1}{10}\, h\, e^{- 2 u^0 - 4 v^0}\, \left[ 4\, \tilde{\xi}_1 + 8\, \tilde{\xi}_2 + \tilde{\xi}_3 - 32\, \tilde{\xi}_4 - 80\, e^{u^0 + 4 v^0}\, \tilde{\phi}_1 - 80\, e^{3 u^0 + 2 v^0}\, \tilde{\phi}_2 \right] \, ,\nonumber\\
\end{align}

\begin{align}\label{phi5}
\tilde{\phi}_{5}^{\prime} =&\, \frac{g_s^{1/2}}{4\, m^2}\, h\, e^{3 z^0 / 2 - 3 \Phi^0 / 4}\,  \Big[ \ell^6 \, e^{15 z^0 / 2 + \Phi^0 / 4}\, \tilde{\xi}_5 \nonumber\\ & + g_s^{1/4}\, m^2\, \left( 4\, \tilde{\phi}_6 - \left( g_2^0 + g_3^0 \right)\, \left[ 8\, \tilde{\phi}_1 - \tilde{\phi}_3 \right] \right) \Big] \, , \nonumber\\
\end{align}

\begin{align}\label{phi6}
\tilde{\phi}_{6}^{\prime} =&\, \frac{1}{2\, g_s\, m^2}\, h\, e^{-2 u^0 - 4 v^0 - 3 z^0 / 2 + 3 \Phi^0 / 4}\, \Big[ \ell^6\, e^{15 z^0 / 2 + \Phi^0 / 4}\, \left( 2 \, e^{4 u^0}\, \tilde{\xi}_6 + e^{4 v^0}\, \tilde{\xi}_7 \right) \nonumber\\ & + 2\, g_s^{1/4}\, m^2\, e^{4 u^0}\, \left[ 4\, \tilde{\phi}_5 +\left( c_2 + g_1^0 \right) \, \left( 8\, \tilde{\phi}_1 + 8\, \tilde{\phi}_2 - \tilde{\phi}_3 \right) \right] \nonumber\\ & + g_s^{1/4}\, m^2\, e^{4 v^0}\, \left( 4\, \tilde{\phi}_5 - \left( c_3 + g_1^0 \right) \tilde{\phi}_3 \right) \Big] \, ,\nonumber\\
\end{align}

\begin{align}\label{phi7}
\tilde{\phi}_{7}^{\prime} =&\, \frac{1}{2\, g_s\, m^2}\, h\, e^{-2 u^0 - 4 v^0 - 3 z^0 / 2 + 3 \Phi^0 / 4}\, \Big[ \ell^6 \, e^{15 z^0 / 2 + \Phi^0 / 4} \, \left( 2 \, e^{4 u^0}\, \tilde{\xi}_6 - e^{4 v^0}\, \tilde{\xi}_7 \right) \nonumber\\ & + 2\, g_s^{1/4}\, m^2\, e^{4 u^0}\, \left[ 4\, \tilde{\phi}_5 +\left( c_2 + g_1^0 \right)\, \left( 8\, \tilde{\phi}_1 + 8\, \tilde{\phi}_2 - \tilde{\phi}_3 \right) \right] \nonumber\\ & - g_s^{1/4}\, m^2\, e^{4 v^0}\, \left( 4\, \tilde{\phi}_5 - \left( c_3 + g_1^0 \right)\, \tilde{\phi}_3 \right) \Big] \, , \nonumber\\
\end{align}

\begin{align}\label{phi4}
\tilde{\phi}_{4}^{\prime} =&\, - \frac{1}{10\, \ell^6}\, h\, e^{- 2 u^0 - 4 v^0 - 15 z^0 / 2 - \Phi^0 / 4}\, \Big[ \ell^6\, e^{15 z^0 / 2 + \Phi^0 /4}\, \left( 8\, \tilde{\xi}_1 - 4\, \tilde{\xi}_2 - 5\, \tilde{\xi}_3 \right) \nonumber\\ & + 80\, \ell^6\, e^{u^0 + 4 v^0 + 15 z^0 / 2 + \Phi^0 / 4}\, \tilde{\phi}_1 - 80\, \ell^6\, e^{3 u^0 + 2 v^0 + 15 z^0 / 2 + \Phi^0 / 4}\, \tilde{\phi}_2 \nonumber\\ & + 40\, g_s^{1/4}\, m^2\, \Big( 4\, \left( g_2^0 + g_3^0 \right)\, \tilde{\phi}_5 + 2\, \left( 2\, g_1^0 + c_2 + c_3 \right)\, \tilde{\phi}_6 + 2\, \left( c_2 - c_3 \right)\, \tilde{\phi}_7 \nonumber\\ & - \left( g_2^0\, \left( c_2 + g_1^0 \right) + g_3^0\, \left( c_3 + g_1^0 \right) \right)\, \tilde{\phi}_4 \Big) \Big] \, .
\end{align}

\section{The force on a probe D2--brane}\label{SF}

In this section we evaluate the force felt by a D2--brane probing a generic linearized deformation of the CGLP background. At first glance, the expression for that force might seem quite involved. Yet, we will show that using the first--order equations of motion most of the terms cancel and the final expression is quite simple, involving only a single mode. This is as expected from previous work on the linearized perturbations around IIB and 11--dimensional BPS solutions~\cite{BGrH,BGGH,BGiH}.

Let's then expose the analytic expression we found for the force exerted on a D2--brane surveying a generic deformation of the supersymmetric CGLP background.

We choose a static gauge for a brane spanning Minkowski space directions, without any gauge field on its world--volume. The DBI Lagrangian reduces to 
\beq
   \mathcal{L}_{DBI}=-V^{DBI}=-T_p\, e^{-\Phi/4}\, g_s^{-3/4}\, \sqrt{-g_{00}\, g_{11}\, g_{22}} = - T_p\, e^{-\Phi/4-15z/2}\, g_s^{-3/4}\,.
\eeq

The only non-zero RR potential is $C_{MNP}$, and the part which gives non--vanishing contribution is given by 
\beq
C_3&=\frac{1}{g_s}\, {\cal K}(r)\, dx^{0}\w dx^{1}\w dx^{2}\, , \qquad \frac{d {\cal K}(r)}{dr}=-K(r) \, ,
\eeq
with $K(r)$ given in equation~\eqref{Kvalue}. The Wess-Zumino piece of the D2--brane action thus reduces to
\beq
  \mathcal{L}_{WZ}=-V^{WZ}=T_p\frac{1}{3!}\varepsilon^{i_1i_2i_3}(C_3)_{i_1i_2i_3}=-T_p\frac{1}{g_s} {\cal K}(r)\,.
\eeq
We can now compute the force on a probe D2--brane (from now on we fix $T_p=1$). At zeroth order we have
\begin{align}
	F^{(0)\,DBI}&=g_s^{-1/2}H'_0e^{-\Phi^0/2-15z^0}=-\frac{4m^2}{\ell^6}g_s^{-1/2}e^{-\Phi^0/2-15z^0-2u^0-4v^0}h\,\left[c_2g^0_2+c_3g^0_3+g_1^0(g_2^0+g_3^0)\right]\nonumber\\
	F^{(0)\,WZ}&=\frac{1}{g_s}K(r)=\frac{4m^2}{\ell^6}g_s^{-1/2}e^{-\Phi^0/2-15z^0-2u^0-4v^0}h\,\left[c_2g^0_2+c_3g^0_3+g_1^0(g_2^0+g_3^0)\right]\nn
\end{align}
and, as further confirmation that everything is under control so far, the two contributions compensate each other, as they should.

As for the first--order contribution to the force, it arises from
\begin{align}
	F^{(1)\,DBI}&=-F^{(0)\, DBI}\left(\frac{1}{4}\phi_4-\frac{15}{2}\phi_3\right)+g_s^{-3/4}\left(\frac{1}{4}\phi'_4+\frac{15}{2}\phi'_3\right)e^{-\frac{\Phi^0}{4}-\frac{15}{2}z^0}\nonumber\\
	F^{(1)\,WZ}&=-F^{(0)\,WZ}\left(\frac{1}{2}\phi_4+\frac{15}{2}\phi_3 -2\phi_1-4\phi_2\right)+g_s^{-3/4}\left(\frac{1}{4}\phi'_4+\frac{15}{2}\phi'_3\right)e^{-\frac{\Phi^0}{4}-\frac{15}{2}z^0}\nonumber\\
		        &\qquad+\frac{4m^2}{\ell^6}g_s^{-1/2}h\,e^{-\frac{\Phi^0}{2}-15 z^0-2u^0-4v^0}\left[c_2\phi_6+c_3\phi_7+\phi_5(g^0_2+g^0_3)+g^0_1(\phi_6+\phi_7)\right]\nn\,.
\end{align}
From these expressions, using the first--order equations~\eqref{phi3},~\eqref{phi4} for $\phi_3$ and $\phi_4$, as it happens, most of the terms at first--order cancel so that the force on a probe D2--brane reduces to 
\begin{align}
  F(r)&=F^{(1)\,DBI}+F^{(1)\,WZ}\nonumber\\
        &=\frac{1}{8g_s^{3/4}}h\,e^{-2u^0-4v^0-\frac{15}{2}z^0-\frac{1}{4}\Phi^0}\txi_3\nonumber\\
        &=\frac{2}{g_s}\frac{X_3\,e^{-8z^0(1)}}{(r^4-1)^{3/2}} \, , \label{force}
\end{align}
where we have made preemptive use of the analytic solution for the mode $\txi_3$, eq.~\eqref{warpmode}, which will be derived in the next section. As an aside, the derivative of the Green's function for the CGLP background~\eqref{GD2} matches the behavior of the force~\eqref{force} (see~\cite{BGGH} for comments on this point). Indeed, allowing only for a dependence on the radial variable, the solution to \beq
\Box \, G=0
\eeq 
evaluated on the CGLP background is
\beq
G(r) = c_1 + c_2\, \left(\frac{r}{\sqrt{r^4-1}}- \eF \, \right)\label{GD2}\, .
\eeq

\section{Prelims: boundary conditions for BPS D2 branes}\label{SBPS}

In this section, as a matter of exposing our method before we focus on the candidate backreaction by anti--D2 branes, we first derive the boundary conditions which correspond to the modes sourced by a stack of branes placed at the tip of the cone.\\

 Let us then consider a set of $N$ ordinary extremal D2 branes smeared on the $S^4$ at the bottom of the throat. For the CGLP background, we can explicitly evaluate the Maxwell charge 
\beq\label{QMAX}
\qqq^{Max}_{CGLP}(r)=\frac{1}{(2\pi\sqrt{\alpha'})^5}\int_{M_6}e^{\Phi/2}*F_4=\frac{4m^2g_s^{-1/2}}{\ell(2\pi\sqrt{\alpha'})^5}\vol(M_6)[g_1(g_2+g_3)+c_2 g_2+c_3 g_3]\,.
\eeq
This quantity exhibits the following zeroth--order IR behavior:
\beq
\qqq_{CGLP}^{IR}=0 \, ,
\eeq
as can be seen from
\beq
g_1^0(g_2^0+g_3^0)+c_2g_2^0+c_3g_3^0 \simeq \frac{7}{128} (r-1)^{3/2}-\frac{77}{512}
   (r-1)^{5/2}+{\cal O}\left((r-1)^{7/2}\right)\, ,
\eeq
using equation~\eqref{c2c3}.\\ 

Within the Ansatz we have been considering, a BPS solution describing the addition of $N$ ordinary BPS D2 branes smeared on the $S^{4}$ in the IR can be found by shifting $g_2$ and $g_3$ such that the combination $g_2 + g_3$ --- which is multiplied by $g_1$ in~\eqref{QMAX} --- does not change:
\beq
g_2^0\rightarrow g_2^0+\frac{32N}{3}\,, \qquad g_3^0\rightarrow g_3^0-\frac{32N}{3}\,.\label{gshift}
\eeq
This way, the charge is shifted as
\beq
\qqq^{Max}_{CGLP}\rightarrow \qqq^{Max}_{CGLP}+ \Delta  \qqq^{Max}_{D2}\, ,
\eeq
with
\beq
\Delta  \qqq^{Max}_{D2}=\frac{4Nm^2}{(2\pi\sqrt{\alpha'})^5}\frac{g_s^{-1/2}}{\ell}\vol(M_6)\,.
\eeq
Note that the flux through $S^4$,
\beq
q_{S^4}=\frac{1}{(2\pi\sqrt{\alpha'})^3}\int_{S^4} F_4=\frac{4mg_s^{-1}}{(2\pi\sqrt{\alpha'})^3}(g_1+c_2)\vol(S^4)\, ,
\eeq
stays unchanged under the shifts~\eqref{gshift}, while the warp factor shifts as
\beq
 H_{0}(r)\rightarrow-\frac{4m^2}{\ell^6}\int^{r}h\,e^{-2u^0-4v^0}\left[g^0_1(g^0_2+g^0_3)+c_2g^0_2+c_3g^0_3+ N\right]d y\, 
\eeq
and now is endowed with a singularity of the kind
\beq
 H(r)	\thicksim\frac{\Delta\qqq_{D2}}{\sqrt{r-1}}\label{warpdiv} \, .
\eeq
This is the expected behavior of the harmonic function for Dp branes smeared on an $S^{r}$ within an otherwise ten--dimensional flat space, which indeed behaves as $\frac{1}{\tau^{7-p-r}}$, where $p=2$ and $r=4$ for the CGLP background.\\ 

Let us now see in more detail how this BPS solution can be reproduced by the first--order perturbation apparatus. First of all, we set to zero all the modes related to supersymmetry--breaking, namely we impose that all the constants $X_a$ and $B _1\thicksim X_3$~\eqref{B1 def}, which enter upon integrating of $\txi$ equations, should vanish. 

Furthermore, the zeroth--order combinations $e^{2u^0}$ and $e^{2v^0}$ reach constant or zero value in the IR ; since we expect that the geometry of the transverse space is not affected by the addition of BPS D2 branes we impose the perturbations associated to $u$ and $v$ to vanish as well. This fixes
\begin{align}
 Y_1^{IR}&= Y_2^{IR}=0\label{bc1} .
\end{align}

In addition, non--singularity of $\phi_5$ and $\phi_7$ (we recall that they enter the fluxes of our Ansatz) is ensured by
\begin{align}
 Y_5^{IR}= -\frac{1}{840} K(-1) (7 Y_3^{IR}+80 Y_6^{IR})\label{bc2}.
\end{align}
The mode $\phi_6$ is regular, and in view of the first--order contribution to \eqref{QMAX}
\beq
\left( g_1^{0} + c_2 \right)\, \phi_6 + \left( g_1^{0} + c_3 \right)\, \phi_7 + \left( g_2^{0} + g_3^{0} \right)\, \phi_5 \simeq - \frac{\sqrt{r-1}}{8}\, \phi_5 + \left( \frac{3}{32} - \frac{r-1}{16}\right)\, \phi_6 - \frac{r-1}{16}\, \phi_7 \, ,
\eeq
one should impose that $\phi_6(r \rightarrow 1)$ be proportional to the number $N$ of BPS D2 branes spread over $S^4$ at the tip.\\

To recap, the above choices of integration constants~\eqref{bc1}--\eqref{bc2} yield the expected behavior for BPS D2 branes added in the supersymmetric CGLP background:
\begin{align}
\phi_1&=0\,, & \phi_2&=0\,,&
\phi_5&={\cal O}\left(r-1\right) \,,
\end{align}
\begin{align}
\phi_3&=-\frac{2Y_7^{IR}}{4-K(-1)^2}\frac{1}{\sqrt{r-1}}+{\cal O}\Big((r-1)^{1/2}\Big)\,, & \phi_4&=-\frac{4Y_7^{IR}}{4-K(-1)^2}\frac{1}{\sqrt{r-1}}+{\cal O}\Big((r-1)^{1/2}\Big)\,,\nn\\
\phi_6&=\frac{1}{2}Y_7^{IR}+{\cal O}\Big((r-1)^{1/2}\Big)\,, & \phi_7&=-\frac{1}{2}Y_7^{IR}+{\cal O}\Big((r-1)^{1/2}\Big)\nn\, .
\end{align}
We recall that $\phi_{1,2}$ denote perturbations of the stretching functions, $\phi_{3,4}$ label perturbations of the warp factor and dilaton, whilst $\phi_{5,6,7}$ are the modes corresponding to the linearized perturbations of the NSNS and RR fluxes of this IIA background.

The integration constant $Y_7^{IR}$ is the only remaining one and is related to the number $N$ of added BPS D2 branes: indeed, the equations for $\phi_6$ and $\phi_7$ reproduce the shift \eqref{gshift}. The warp factor, along with the dilaton, acquires the expected singularity and
\beq
H=e^{8z_0}\, \left( 1+8\, \phi_3 \right)\, , \, \, \, e^{\Phi}=e^{\Phi_0}\, \left(1 + \phi_4\right)=e^{2 z_0}\, (1 + 2\, \phi_3) \, ,
\eeq
in accordance with $e^{\Phi} \thicksim H^{1/4}$.

\section{Assessing the anti--D2 brane solution}\label{SAD2}

The final step and main aim of our analysis is to determine how, within the space of generic linearized deformations of the IIA CGLP background, one can account for the backreaction due to the addition of anti--D2 branes smeared on the $S^4$ at the tip of the warped throat.\\

As the prime physical requirement we should impose that the force felt by a D2 brane probing the backreaction due to this stack of anti--D2 branes be non-vanishing. So, we are forbidden from turning off the corresponding mode which appears in the expression~\eqref{force} of the force, and enters the various expressions for the modes $\phi_a$ by means of the shorthand combination 
\beq
B_1=\frac{m^2}{\ell^6}X_3e^{-8z_0(1)}\, .
\eeq 

As our next set of IR boundary conditions, let us recall that the modes $\phi_3$ and $\phi_4$ associated to the perturbation of the warp factor and the dilaton must exhibit no worse than a $1/\sqrt{r-1} \sim 1/\tau$ behavior (cf. equation~\eqref{tau to r}). Such a behavior is in accordance with the Coulomb--like divergence associated to anti--D2 branes smeared over the $S^4$ at the tip of the warped throat. 

Inspecting the IR expansions of the deformation modes $\phi_a$, every piece that is more singular than the aforementioned $1/\sqrt{r-1}$ behavior will be culled by tuning appropriate combinations of the $X$'s and the $Y$'s integration constants parametrizing the space of generic linearized perturbations of the CGLP background.\\ 

Another, equivalent but slightly less liberal, criterion that we are about to consider focuses on allowing or discarding various pieces from the $\phi_a$'s IR expansions depending on their contribution to the energy. More precisely, we consider the kinetic energy~\eqref{kinetic} and the potential energy~\eqref{potential} obtained by reducing the IIA supergravity Ansatz~\eqref{Ansatz} to a one--dimensional sigma model. 

For instance, the energy associated to the first--order perturbation of the dilaton and warp factor is obtained by expanding to second--order the corresponding terms from~\eqref{kinetic}:
\begin{align}
& \frac{e^{2\, \left( u^0 + \phi_1 \right) + 4\, \left( v^0 + \phi_2 \right)}}{h}\, \Big[ -30\, \left(z^{0\,\prime} + \phi_3^{\prime} \right)^2 - \frac{1}{2} \left( \Phi^{0\,\prime} + \phi_4^{\prime} \right)^2  \Big] \nonumber\\ &
\leadsto \nonumber\\ &
\frac{e^{2\, u^0 + 4\, v^0}}{h}\, \Big[ -30\, \phi_3^{\prime \, 2} - \frac{1}{2}\, \phi_4^{\prime\, 2} - 2\, \left( \phi_1 + 2\, \phi_2 \right)\, \left( \Phi^{0\,\prime}\, \phi_4^{\prime} + 60\, z^{0\,\prime}\, \phi_3^{\prime} \right) \Big] 
\end{align}
The energy associated to the deformation of the warp factor and dilaton exhibits the following singular behavior
\beq
\left( r-1 \right)^{3/2}\, \left( \frac{d\phi_{3,4}}{dr} \right)^2 \sim \frac{1}{(r-1)^{3/2}}\nn\,\label{thresh} \, ,
\eeq
where as a matter of course we neglect less diverging terms. This behavior sets the threshold for what we consider an allowable singularity in the energy.\\

Note that, as it turns out, for all practical purposes we can neglect contributions of the type $\phi_{a} \phi_{b}$ and $\phi_{a}^{\prime}\, \phi_{b}$ for $a \neq b$: they only contribute to sub--leading divergences. In addition, there is no contribution to the energy that is first--order in the SUSY--breaking parameters, since we are expanding around a saddle point.\\

 Another remark is in order. We have considered linearized deformation for the fields entering the supergravity Ansatz~\eqref{Ansatz}, namely we have expanded as
 \beq
 \phi_a = \phi^{0}_{a} + \phi^{1}_{a}(X,Y) \, ,
 \eeq 
 with $X_{i}$ and $Y_{i}$ being implicitly the small supersymmetry--breaking expansion parameters. On the other hand, we are considering quadratic contributions of the $\phi^{1}_{a}$'s to the energy. 
 
The reason why we do not stop at first--order contributions to the energy from those deformation modes is that we have expanded around a saddle point. Had we gone as far as computing 2nd order expansions of the deformation modes, namely
 \beq
 \phi_a = \phi^{0}_{a} + \phi^{1}_{a}(X,Y) + \phi^{2}_{a}(X,Y, Z,W) \, , 
 \eeq
which is an achievable if strenuous task, it might well happen that the singularities we are about to expose might cancel against truly second order contributions to the energy. By this we mean contributions of the type $ \phi^{2}_{a} \phi^{0}_{b}$, in addition to those of the form $\left( \phi^{1}_{a} \right)^2 \, \phi^{0}_{b}$ that we presently consider.\\ 

Everything is now in place to show that the candidate IIA supergravity dual to metastable supersymmetry--breaking that would be obtained out of backreacting $\overline{D_2}$'s spread over the $S^4$ in the far IR of the CGLP background comes with an irretrievable IR singularity. Indeed, we are going to show that it is not possible to simultaneously satisfy the two previously mentioned physical requirements. 

In point of fact, there is a singularity associated to the NSNS and RR fluxes that is worse than the ones we allow, namely those that are physical and should be kept based on their identification with the effect of adding anti--D2 branes to uplift the AdS minimum of the potential. There is only one way of getting rid of that ``unphysical'' singularity: it entails setting to zero the single mode entering the force felt by a brane probing the non--supersymmetric backreaction by $\overline{D_2}$'s. So, our two sensible IR boundary conditions are incompatible.

Ensuring that there is a force exerted on a probe D2--brane by the anti--D2's at the tip results in a $\frac{1}{\left( r-1 \right)^3} \sim \frac{1}{\tau^6}$ singular contribution to the energy, stemming from the NSNS or the RR field strength. Such a singularity is worse than the ones it is sensible to a priori allow, namely $\frac{1}{(r-1)^{3/2}}$ singularities or milder ones, associated to the smeared $\overline{D_2}$'s.\\

Let us see how this comes about with full details. First of all, note that the potentially most divergent deformation modes is $\phi_7$: its IR series expansion~\eqref{phi7IRsol} displays $\frac{1}{r-1}$ and $\frac{\log(r-1)}{r-1}$ pieces. That mode, $\phi_7$, contributes only to the deformation of the NSNS 3--form field strength
\beq
\ell \, \delta H_3 = m\, \left[ \left( \phi_6 + \phi_7 \right)\, U_3 + \phi_6^{\prime}\, dr\w U_2 + \phi_7^{\prime}\, dr\w J_2 \right] \, .
\eeq
In view of~\eqref{kinetic} and~\eqref{potential}, the leading contribution to the energy from the deformation of the NSNS 3--form is
\beq\label{H3 sing energy}
- \frac{m^2}{2\, \ell^6}\, \frac{e^{2\, u^0 + 4\, v^0 - 8\, z^0}}{h}\, \left[ \phi_6^{\prime \, 2}\, e^{-4\, u^0} + 2\, \phi_{7}^{\prime \, 2} \, e^{-4\, v^0} \right] - 2\, \frac{m^2}{\ell^6}\, h\, e^{-8\, z^0}\, \left[ \phi_6 + \phi_7 \right]^2 \, .
\eeq
There is another potential contribution from~\eqref{potential} which involves $\phi_6$ and $\phi_7$. It is easily seen that it is sub--leading. Now, what is the IR singular behavior of~\eqref{H3 sing energy} ?  We focus on the most singular piece of $\phi_7 \sim \frac{1}{r-1}$ and its derivative. It entails the following singular behavior
\beq
- \frac{m^2}{\ell^6}\, e^{-8\, z^0(r)}\, \left[ \frac{e^{2\, u^0(r)}}{h(r)}\, \left( \frac{d}{dr}\, \frac{1}{(r-1)} \right)^2 + 2\, h(r)\, \left( \frac{1}{(r-1)} \right)^2 \right] \sim \frac{1}{(r-1)^{5/2}} \, .
\eeq

According to our physical criterion pertaining to the energy, we should then discard the most IR--divergent piece of $\phi_7$, see~\eqref{phi7IRsol}. This is achieved by imposing
\beq
X_5= \frac{1}{168}\, \left[ 3\, \left( 17 + K(-1)^2 \right)\, B_1 + 56\, K(-1)\,\left( 3\, X_6 - 2\, X_7 \right)\, \right] \, ,\label{phi7 c1}
\eeq
\begin{align}
X_1 = \frac{1}{86016}\, \Big[ & 6048\, B_1 + 1032192\, X_4 + 215040\, Y_1^{IR} - 2580480\, Y_5^{IR} + 215040\, E(-1)\, Y_2^{IR} \nonumber\\ & + 235200\, K(-1)\, X_6 - 139360\, K(-1)\, X_7 + 133120\, K(-1)\, Y_2^{IR} \nonumber\\ & - 21504\, K(-1)\, Y_3^{IR} - 245760\, K(-1)\, Y_6^{IR} + 8364\, K(-1)^2\, B_1 \nonumber\\ & - 27216\, K(-1)^3\, X_6 +11304\, K(-1)^3\, X_7 - 1809\, K(-1)^4\, B_1 \Big] \, , \label{phi7 c2}
\end{align}
where~\eqref{phi7 c1} has been applied to obtain~\eqref{phi7 c2} out of the combination of $X$'s and $Y$'s from the $\frac{1}{(r-1)}$ part of $\phi_7$'s IR expansion.

We now turn our attention to getting rid of the singularities stemming from the RR flux and $\phi_5$. First of all, note that the condition~\eqref{phi7 c1} washes out, at no extra cost, the leading $\frac{\log(r-1)}{\sqrt{r-1}}$ part of $\phi_5$'s IR asymptotics.

Still, one should enforce that the $\frac{1}{\sqrt{r-1}}$ part of $\phi_5$'s IR expansion be wiped out by appropriately tuning some of the $X$'s and $Y$'s. Indeed, if kept unchecked, that divergent piece would yield a singularity in the energy arising from the RR flux:
\begin{align}
& - 2 \frac{m^2}{\ell^6}\, \frac{e^{-8\, z^0 - 9\, \phi_3 + \phi_4 / 2}}{h}\, \left( g_1^{0\, \prime} + \phi_5^{\prime} \right)^2 \nonumber\\ & - 4\, \frac{m^2}{\ell^6}\, e^{- 8\, z^0 - 9\, \phi_3 + \phi_4 / 2 + 2\, u^0 + 2\, \phi_1}\, h\, \left[ 2\, \left( g_1^{0} + c_2 + \phi_5 \right)^2\, e^{-4\, v^0 - 4\, \phi_2} + \left( g_1^{0} + c_3 + \phi_5 \right)^2\, e^{-4\, u^0 - 4\, \phi_1} \right] \nonumber\\ &
\leadsto \frac{1}{\left( r - 1 \right)^{5/2}} \, ,
\end{align}
which is beyond the energy threshold~\eqref{thresh} and should be culled. To get rid of that singular piece from $\phi_5$, one must exact
\begin{align}\label{phi5 cond}
&  - 32\, K(-1)\, \left( 6384\, X_6 - 3711\, X_7 + 4160\, Y_2^{IR} - 672\, Y_3^{IR} - 7680\, Y_6^{IR} \right)\nn\\ & 6\, K(-1)^2\, \left( 1795\, B_1 - 3976\, X_5 \right)+ 152\, K(-1)^3\, \left( 336\, X_6 - 179\, X_7 \right) \nn\\&  + 2235\, K(-1)^4\, B_1 - 42024\, B_1 + 1344\, \big[ 64\, X_1 - 768\, X_4 - 23\, X_5\nn\\& - 160\, ( Y_1^{IR} + E(-1)\, Y_2^{IR} - 12\, Y_5^{IR} )\, \big] = 0 \, .
\end{align}
We have finally reached the punchline of our analysis: taking into account the conditions~\eqref{phi7 c1}--\eqref{phi7 c2} that did arise from ensuring that no ``unphysical'' singularity pops out of the NSNS flux, it turns out that~\eqref{phi5 cond} yields
\beq
11340\, \left(4 - K(-1)^2 \right)\, B_1 = 0 \, ,
\eeq
in blatant opposition to the physical requirement that a D2--brane probing the non--supersym-\newline{}
metric deformation of the CGLP background experiences a non--vanishing force !\\

We have therefore come to the conclusion that a careful analysis of the backreaction of anti--D2 branes on the CGLP background inevitably results in an IR singularity. By focusing on two particular flux elements for which the energy contribution can be easily calculated, we have shown that it is not possible to avoid a singular behavior provided we want to keep the $B_1$ mode entering the expression for the force~\eqref{force} to be non--vanishing. 

One has to face that at least one of the perturbed NSNS or RR fluxes contributes to a divergent energy density and to a divergent action as well (given that the factor $\sqrt{g_{10}}\simeq\sqrt{r-1}$ appearing in the ten--dimensional action~\eqref{SUGRAIIaEinsteinFrame} is not enough to make the action finite in the IR), much as is the case in~\cite{BGiH}. The key difference from~\cite{BGiH} lies in the fact that in our case the singular behavior is not at all sub--leading.\\ 

The above type IIA analysis completes the program of investigating the would--be backreacted supergravity duals to metastable supersymmetry--breaking vacua, which was originally started in a type IIB setting~\cite{BGrH}, and next considered in~\cite{BGiH} in an 11--dimensional context.
It would be of much interest to consider other backgrounds and/or, as explained at the beginning of this Section, to go to higher--order in the perturbations around those BPS solutions. It might be that an absence of the nasty singularities we have kept on encountering so far could be used in order to discriminate among solutions of the landscape string theory vacua.

\vspace{1cm}
 \noindent {\bf Acknowledgements}:\\
 \noindent It is a pleasure to thank Iosif Bena, Tae--Joon Cho, Anatoly Dymarsky, Hadi Godazgar, Mahdi Godazgar, Mariana Gra\~na, Anshuman Maharana, Stefano Massai, Piljin Yi and Thomas Van Riet for discussions and interest in this work. G.~G.~is grateful to the DAMTP at Cambridge University and the Simons Center for Geometry and Physics for hospitality while this paper was being completed. This work was supported in part by a Contrat de Formation par la Recherche of CEA/Saclay.

\begin{appendix}

\section{The space of linearized deformations}\label{Sspace}

\subsection{The $\txi_a$ system}

\subsubsection{Structure of the solutions}

We start with some remarks on our approach to determining solutions to the set of equations~\eqref{xit3}--\eqref{xit2}. In the case at hand, we were able to find fully analytic expressions\footnote{It is important to have a solution expressed in terms of the least possible number of nested integrals. As happens in previous similar work~\cite{BGrH, BGiH, BGGHM1}, it is usually not possible to find a fully integrated solution and one then has to be content with series expansion ; if the number of nested integrals is important, that quickly becomes burdensome. When counting of nested integrals we do not take into account the one which enters the definition of the elliptic functions.} for all the $\tilde{\xi_a}$. We present the comments in the order in which the corresponding equations have to be solved.

The first equation we have to solve is the one for $\tilde{\xi}_3$. Upon recognizing the algebraic expression for $K(r)$~\eqref{Kvalue} and keeping in mind that for a BPS solution the following identity holds
\beq
K_{(0)}(r) = - \frac{H_{(0)}^{\prime}}{H_{(0)}(r)^{2}} \, , 
\eeq
it can be expressed as
 \beq
\txi'_3=\frac{H'_0}{H_0}\,\txi_3\label{warpmode}
\eeq

The solution obviously is:
\beq
\txi_3(r)=X_3 H_0(r)e^{-8z^0(1)}\, .
\eeq 
Let us introduce at this stage the constant $B_1$, which we find convenient to use in order to avoid extra clutter
\beq\label{B1 def}
B_1=\frac{m^2}{\ell^6}X_3e^{-8z^0(1)} \, .
\eeq

The next step is to explicitly perform the integration entering the expression for the CGLP background warp factor $H_0(r)$, which we rewrite here as
\beq
H_0(r)=\frac{m^2}{\ell^6}\int_r^{\infty} \,y^5\left[u_3(y)-u_1(y)\right]u_1(y)\, dy
\eeq
The integrand has the following structure\footnote{We adopt here, and for the remainder of the paper, the following calligraphic notation for the incomplete elliptic integral of the first kind $F$:
\beq
\fff(r) \equiv F(\mbox{arcsin}(1/r),-1))
\eeq 
and, similarly, later on we will refer to
\beq
\eee(r) \equiv E(\mbox{arcsin}(1/r),-1))
\eeq
as the incomplete elliptic integral of the second kind $E$. Cf.~also Appendix \ref{ellfunc}.}
\beq
\alpha_2\, \fff(r)^2 + \alpha_1\, \fff(r) + \alpha_0 \, ,
\eeq
with $\alpha_i$ some functions of $r$ which do not involve $\fff$. We simply apply integration by parts (in the following we drop the radial dependence for ease of notation):
\begin{align}
\int \alpha_2\fff^2+\alpha_1\fff+\alpha_0&=A_2 \fff^2+\int\left(\alpha_1-2\fff'A_2\right)+\int\alpha_0\nonumber\\
							  &=A_2\fff^2+A_3\fff+\int(\alpha_0-\fff'A_3)\nonumber\\
							  &=A_2\fff^2+A_3\fff+A_4 	\, ,
\end{align}
where the labels introduced above denote the following:
\begin{align}
\fff'&=\frac{d}{d y}F(\arcsin(1/y)|-1)=-\frac{1}{\sqrt{y^4-1}}\, , \nonumber\\
\alpha_3 &=\alpha_1-2\fff'A_2\, , \nonumber\\
\alpha_4 &=\alpha_0-\fff'A_3\, , \nonumber\\
A_i &=\int\alpha_i\,.
\end{align}
Once we have a primitive we have just to evaluate it at the two extrema of integration to get an analytic expression for $H_0$, and therefore for $\tilde{\xi}_3$.

The equations for $\txi_7$ is:
\beq\label{xit7eq}
\txi'_7=-\frac{3}{64}\frac{m^2}{l^6}h\,e^{-2u^0-4v^0}H^{-1}_0\txi_3 =-\frac{3}{4}\, \frac{B_1}{(r^4-1)^{3/2}}
\eeq
which can be directly integrated.

The functions $\txi_5$ and $\txi_6$ are coupled into a subsystem of ODE's, which we can rewrite as
\begin{align}
\txi'_5&=-2h\,(2e^{2u^0-4v^0}+e^{-2u^0})\,\txi_6 - 2h\,e^{-2u^0}\,\txi_7 -\frac{32}{3}f_1\,\txi'_7\label{xi561}\\
\txi'_6&=-h\,\txi_5-\frac{8}{3}\frac{1}{h}e^{-2u^0+4v^0}f_2\,\txi'_7\label{xi562}
\end{align}
In order to obtain a solution, we first have to solve for the homogeneous system ; we  arrange the two basis vectors of the space of homogeneous solutions in the so--called fundamental matrix
\beq
\tilde{\Xi}_{\scriptscriptstyle{56}}=\left(\begin{array}{c|c}
					\frac{(3r^4-1)}{r^4(r^4-1)} & \frac{r(6r^8-6r^4-1)}{r^3\sqrt{r^4-1}}-\frac{3r^4-1}{r^4(r^4-1)}\fff(r) \\ 
					\hline
					\frac{1}{r\sqrt{r^4-1}} & 1-\frac{3r^4}{2}-\frac{1}{r\sqrt{r^4-1}} \fff(r)
	           	\end{array}\right) \,.
\eeq
The solution to the inhomogeneous system is then expressed as
\beq
\left(\begin{array}{c} \tilde{\xi}_5(r)\\ \tilde{\xi}_6(r)\end{array}\right)=\tilde{\Xi}_{\scriptscriptstyle{56}}(r)X_{\scriptscriptstyle{56}}+\tilde{\Xi}(r)\int^r\,\tilde{\Xi}_{\scriptscriptstyle{56}}(y)^{-1}g^{\xi}_{\scriptscriptstyle{56}}(y)\,d y\nn
\eeq
where $X_{\scriptscriptstyle{56}}=(X_5,X_6)$ are integration constants, and $g^{\xi}_{\scriptscriptstyle{56}}=(g^{\xi}_{\scriptscriptstyle{5}},g^{\xi}_{\scriptscriptstyle{6}})$ is a book--keeping for the non-homogeneous terms entering equations~\eqref{xi561} and \eqref{xi562}.

The equation for $\txi_4$ is entirely non-homogeneous and depends on $\txi_5$ and $\txi_6$. We can rewrite it as follows:
\beq
 \txi'_4=\frac{3}{4}h\,f_1\,\txi_5-\frac{3}{4}(f_2+f_3)\,\txi_6-\frac{3}{4}\,f_3\txi_7-\frac{B_1}{32}h\,e^{u^0}(2u_3-3)u_1\,.
\eeq
which we managed to integrate.

Finally, the functions $\txi_1$ and $\txi_2$ are entangled into the following system of first--order differential equations:
\begin{align}\label{eq:xit12eqs}
\txi'_1&= h\,e^{-u^0}\,\txi_1 + h\,e^{u^0-2v^0}\,\txi_2 -2(f_2-f_3)\txi_6+2f_3\txi_7-\frac{B_1}{8}r(2u_3-3)u_1\,,\\
\txi'_2&=h\,e^{-u^0}\,\txi_1+3h\,e^{u^0-2v^0}\,\txi_2-2(3f_2-f_3)\,\txi_6+2f_3\,\txi_7+\frac{B_1}{8}r(2u_3-3)u_1\,.
\end{align}
whose fundamental matrix $\tilde{\Xi}_{\scriptscriptstyle{12}}$ reads
\beq
\tilde{\Xi}_{\scriptscriptstyle{12}}=\left(\begin{array}{c|c}
					r^4-1 & \frac{\sqrt{r^4-1}}{r}\left(1-r\sqrt{r^4-1}(\eee(r)-\fff(r)\right)\\ \hline
					2r^4 & 	-2r^4\left(\eee(r)-\fff(r)\right)
					\end{array} \right)\,.
\eeq
Analytic expressions for $\txi_{1,2}$ are listed in the next subsection, along with solutions for their siblings.

\subsubsection{Fully analytic expressions for the $\xi^a$ modes}

Here, we collate analytic solutions we derived for the $\tilde{\xi}$ system\footnote{We made sure that those solutions are explictly real, which straightforward if gruelling successive integration by parts do not immediately yield.}

\begin{align}\label{eq:xit1sol}
 \txi_1&=\fff(r)^3\left(-B_1\frac{r^4+1}{112r^5(r^4-1)^{3/2}}\right)\nonumber\\
           &+\fff(r)^2\left(B_1\frac{189r^{12}-258r^8+r^4+48}{1792r^4(r^4-1)}+(45B_1K(-1)-168X_6+112X_7)\frac{r^4+1}{2688r^5(r^4-1)^{3/2}}\right)\nonumber\\
           &+\fff(r)\Bigg(-B_1\frac{69r^{12}-114r^8+61r^4-24}{896r^3(r^4-1)^{3/2}}-B_1K(-1)\frac{315r^{12}-390r^8-53r^4+120}{3584r^4(r^4-1)}\nonumber\\
           &\qquad\qquad+X_2(r^4-1)-X_6\frac{63r^{12}-78r^8+31r^4-8}{64r^4(r^4-1)}-X_7\frac{9r^{12}-18r^8-7r^4+8}{96r^4(r^4-1)}\nonumber\\
           &\qquad\qquad+(24X_5+K(-1)(24X_6-16X_7-3B_1K(-1)))\frac{r^4+1}{384r^5(r^4-1)^{3/2}}\Bigg)\nonumber\\
           &\,\,\,\,-B_1\frac{51r^8-75r^4+16}{1792r^2(r^4-1)}+B_1K(-1)\frac{315r^{12}-516r^8+229r^4-60}{3584r^3(r^4-1)^{3/2}}+X_1(r^4-1)\nonumber\\
           &\,\,\,\,-B_1K(-1)^2\frac{63r^{12}-126r^8+63r^4-4}{512r^4(r^4-1)}+X_2\frac{\sqrt{r^4-1}}{r}-X_2(r^4-1)\eee(r)+X_5\frac{2r^4-1}{16r^4(r^4-1)}\nonumber\\
           &\,\,\,\,-X_5K(-1)\frac{r^4+1}{16r^5(r^4-1)^{3/2}}-X_6\frac{33r^8-35r^4+4}{64r^3\sqrt{r^4-1}}+X_6K(-1)\frac{63r^{12}-78r^8+23r^4-4}{64r^4(r^4-1)}\nonumber\\
           &\,\,\,\,+X_7\frac{9r^8-11r^4+4}{96r^3\sqrt{r^4-1}}+X_7K(-1)\frac{9r^{12}-18r^8+r^4+4}{96r^4(r^4-1)}\, ,
\end{align} \\\\\\\\\\\\\\\\\\\\\\\
\begin{align}\label{eq:xit2sol}
 \txi_2&=\fff(r)^3\left(B_1\frac{r^4-3}{112r^5(r^4-1)^{3/2}}\right)\nonumber\\
           &+\fff(r)^2\left(B_1\frac{189r^{16}-438r^{12}+241r^8+52r^4-16}{896r^4(r^4-1)^2}-\frac{(45B_1K(-1)-168X_6+112X_7)(r^4-3)}{2688r^5(r^4-1)^{3/2}}\right)\nonumber\\
           &+\fff(r)\Bigg(-B_1\frac{69r^{12}-132r^8+25r^4+20}{448r^3(r^4-1)^{3/2}}-B_1K(-1)\frac{315r^{16}-750r^{12}+427r^8+76r^4+44}{1792r^4(r^4-1)^2}\nonumber\\
           &\qquad\qquad+X_2\, 2r^4-X_6\frac{63r^{12}-87r^8+40r^4-12}{32r^4(r^4-1)}-X_7\frac{9r^{12}-9r^8-16r^4+12}{48r^4(r^4-1)}\nonumber\\
           &\qquad\qquad+(K(-1)(3B_1K(-1)-24X_6+16X_7)-24X_5)\frac{r^4-3}{384r^5(r^4-1)^{3/2}}\Bigg)\nonumber\\
           &\,\,\,\,-B_1\frac{51r^8-30r^4-32}{896r^2(r^4-1)}+B_1K(-1)\frac{315r^{12}-561r^8+40r^4+134}{1792r^3(r^4-1)^{3/2}}+X_1\,2r^4-X_2\,2r^4\eee(r)\nonumber\\
           &\,\,\,\,-B_1K(-1)^2\frac{63r^{16}-126r^{12}+63r^8+2r^4-10}{256r^4(r^4-1)^2}+X_5\frac{4r^4-3}{16r^4(r^4-1)}+X_5K(-1)\frac{r^4-3}{16r^5(r^4-1)^{3/2}}\nonumber\\
           &\,\,\,\,-X_6\frac{33r^8-38r^4+6}{32r^3\sqrt{r^4-1}}+X_6K(-1)\frac{63r^{12}-87r^8+32r^4-6}{32r^4(r^4-1)}\nonumber\\
           &\,\,\,\,+X_7\frac{9r^8-14r^4+6}{48r^3\sqrt{r^4-1}}+X_7K(-1)\frac{9r^{12}-9r^8-8r^4+6}{48r^4(r^4-1)} \, , \\\nonumber\\
\txi_3(r)&=X_3e^{-8z_0(1)}H_0(r) \, ,
\end{align}
where
\begin{align}
H_0(r)&=\frac{m^2}{2\ell^6}\fff(r)^2\left(\frac{3}{32}-\frac{1}{8r^4(r^4-1)^2}\right)\nonumber\\
	   &\,\,\,\,\,\,\,-\frac{m^2}{2l^6}\fff(r)\left(\frac{3r^8+3r^4-4}{16r^3(r^4-1)^{3/2}}+\frac{K(-1)}{16}\left(3-\frac{4}{r^4(r^4-1)^2}\right)\right)\nonumber\\
	   &\,\,\,\,\,\,\,+\frac{m^2}{2l^6}\left(\frac{3r^4-4}{32r^2(r^4-1)}+\frac{3r^8+3r^4-4}{16r^3(r^4-1)^{3/2}}K(-1)-\frac{K(-1)^2}{8r^4(r^4-1)^2}\right)\, ,
\end{align}

\begin{align}\label{eq:xit4sol}
\!\!\!\!\!\!\!\!\! \txi_4&=\fff(r)^3\left(\frac{3B_1(3r^4-1)}{448r^5(r^4-1)^{3/2}}\right)+\nonumber\\
            &+\fff(r)^2\left(\frac{B_1(111r^{12}-222r^8+99r^4-16)}{3584r^4(r^4-1)^2}+\frac{(3r^4-1)}{3584r^5(r^4-1)^{3/2}}(168X_6-112X_7-45B_1K(-1))\right)\nonumber\\
            &+\fff(r)\Bigg(-\frac{B_1(15r^8-12r^4+10)}{896r^3(r^4-1)^{3/2}}-\frac{B_1K(-1)(201r^{12}-402r^8+45r^4+44)}{7168r^4(r^4-1)^2}+\nonumber\\
            &+\frac{3r^4-1}{512r^5(r^4-1)^{3/2}}\left(-24X_5+K(-1)\left(3B_1K(-1)-24X_6+16X_7\right)\right)+\frac{9r^8-9r^4+4}{128r^4(r^4-1)}(3X_6-2X_7)\Bigg)\nonumber\\
            &-\frac{B_1(51r^4-32)}{3584r^2(r^4-1)}+\frac{B_1K(-1)(201r^8-231r^4+134)}{7168r^3(r^4-1)^{3/2}}-\frac{B_1K(-1)^2(9r^4-5)}{512r^4(r^4-1)^2}+\frac{3K(-1)(3r^4-1)}{64r^5(r^4-1)^{3/2}}X_5\nonumber\\
            &+\frac{3r^4-2}{128r^3\sqrt{r^4-1}}(3X_6-2X_7)-\frac{3X_5+K(-1)(3X_6-2X_7)}{64r^4(r^4-1)}+X_4\, ,
\end{align}
\begin{align}
  \txi_5&=\fff(r)^2\left(\frac{B_1(1-3r^4)}{7r^4(r^4-1)}\right) \nonumber\\
            &\,\,\,\,\,\,\,+\fff(r)\left(\frac{B_1K(-1)(3r^4-1)}{8r^4(r^4-1)}-\frac{(3r^4-1)(3X_6-2X_7)}{3r^4(r^4-1)}-\frac{3B_1(5r^8-5r^4-2)}{28r^3\sqrt{r^4-1}}\right)\nonumber\\
            &\,\,\,\,\,\,\,+\frac{B_1(15r^8-21r^4+10)}{28r^2(r^4-1)}-\frac{3B_1K(-1)}{8r^3\sqrt{r^4-1}}+\frac{(3r^4-1)}{r^4(r^4-1)}X_5 + 6r\sqrt{r^4-1}\,X_6-\frac{3X_6-2X_7}{3r^3\sqrt{r^4-1}} \, , \nonumber\\
            & \nonumber\\
  \txi_6&=\fff(r)^2\left(-\frac{B_1}{7r\sqrt{r^4-1}}\right)\nonumber\\
            &\,\,\,\,\,\,\,+\fff(r)\left(\frac{B_1(15r^8+3r^4-4)}{112(r^4-1)}+\frac{B_1K(-1)}{8r\sqrt{r^4-1}}-\frac{3X_6-2X_7}{3r\sqrt{r^4-1}}\right)\nonumber\\
            &\,\,\,\,\,\,\,-\frac{3B_1r(5r^4+4)}{112\sqrt{r^4-1}}-\frac{B_1K(-1)}{8(r^4-1)}+\frac{X_5}{r\sqrt{r^4-1}}+\left(1-\frac{3r^4}{2}\right)X_6-\frac{2}{3}X_7 \, , \\
 \txi_7(r)&=X_7+\frac{3}{8}B_1\left[\frac{r}{\sqrt{r^4-1}}-\fff(r)\right] \, . \label{solxit7}
\end{align}

\subsection{The $\phi_a$ equations}

\subsubsection{The structure of their solutions}

The $\tilde{\phi}$ system does not admit a fully analytic solution. We are thus forced to consider either numerical work as in~\cite{BGGHM2}, or to series expansions, that latter option meeting our the needs of the present paper. Here, we present the equations, and show that we can find solutions up to three nested integrals. This might prove helpful to future work. Once more, the presentation follows the order in which the equations have to be solved. 

We report our final results back again to the $\phi$ basis, as it is convenient to impose boundary conditions singling out the effect of anti--D2 branes in these variables.

The functions $\tphi_1$ and $\tphi_2$ are coupled and the system is
\begin{align}
          \tphi_1'&=-h\,e^{-u^0}\tphi_1-h\,e^{u^0-2v^0}\tphi_2+\frac{1}{20}h\,e^{-2u^0-4v^0}(\txi_1+2\txi_2)\,,\label{p121}\\
          \tphi_2'&=-h\,e^{-u^0}\tphi_1-3h\,e^{u^0-2v^0}\tphi_2+\frac{1}{20}h\,e^{-2u^0-4v^0}(4\txi_1+3\txi_2)\,.\label{p122}
 \end{align}
The corresponding fundamental matrix is
\beq
 \tilde\Upsilon_{\scriptscriptstyle{12}}=\left(\begin{array}{c|c}\
 				\frac{r^4+1}{r^3\sqrt{r^4-1}} &\frac{1}{r^4}+\frac{r^4+1}{r^3\sqrt{r^4-1}}\left(\eee(r)-\fff(r)\right)\\ \hline
				\frac{3-r^4}{r^3\sqrt{r^4-1}} & \frac{3}{r^4}+\frac{3-r^4}{r^3\sqrt{r^4-1}}\left(\eee(r)-\fff(r)\right)
 			\end{array}\right)\,.
\eeq
A formal solution is thus 
\beq\label{formal_phi12}
\begin{pmatrix}
  \tphi_1(r) \\
  \tphi_2(r)
\end{pmatrix}=\tilde\Upsilon_{\scriptscriptstyle{12}}(r)Y_{\scriptscriptstyle{12}} + \tilde\Upsilon_{\scriptscriptstyle{12}}(r)\int^{y} \tilde\Upsilon_{\scriptscriptstyle{12}}^{-1}(y)g_{\scriptscriptstyle{12}}^{\phi}(y)dy\,.
\eeq
where $Y_{\scriptscriptstyle{12}}=(Y_{\scriptscriptstyle{1}},Y_{\scriptscriptstyle{2}})$ are integration constants, and $g_{\scriptscriptstyle{12}}^{\phi}=(g_{\scriptscriptstyle{1}}^{\phi},g_{\scriptscriptstyle{2}}^{\phi})$ encodes the non-homogeneous terms in the couple of equations \eqref{p121}--\eqref{p122} above. Some of the integrals can be explicitly done but sadly there are some terms for which we were unable to find a primitive. We thus have a semi--analytic solution, that is up to an implicit integral.

We can use the following relation arising from the equation for $\tphi_1$,
\beq
 -h\,e^{-u^0}\tphi_1-h\,e^{u^0-2v^0}\tphi_2=\tphi'_1-\frac{h}{20}e^{-2u^0-4v^0}\,,
\eeq 
in order to simplify the equation for $\tphi_3$, which will then take the form
\beq
\tphi'_3=8\tphi'_1+\frac{h}{10}e^{-2u^0-4v^0}\left(\txi_3-32\,\txi_4\right)
\eeq
and has the following solution
\beq
\tphi_3(r)=8\tphi_1(r)+\frac{8}{3}\int^r\frac{\txi_3}{(y^4-1)^{3/2}}\,dy-\frac{256}{5}\int^r\frac{\txi_4}{(y^4-1)^{3/2}}\,dy+Y_3\,,
\eeq
which is again implicitly defined in terms of a single integral.

As for the modes $\tphi_5$ and $\tphi_6$, they are coupled and the relevant sub--system is
\begin{align}
\tphi'_5&=h\,\tphi_6+\frac{\ell^6}{4m^2}h\,H_0\,\txi_5-\frac{h}{4}f_1\left(8\tphi_1-\tphi_3\right) \, , \nonumber\\
\tphi'_6&=2h\,e^{2u^0}\left(2e^{-4v^0}+e^{-4u^0}\right)\tphi_5+\frac{\ell^6}{m^2}h\,H_0 e^{2u^0-4v^0}\txi_6+\frac{\ell^6}{2m^2}e^{-2u^0}h\,H_0\,\txi_7\nonumber\\
		&\qquad+\frac{f_2}{4}\left(8\tphi_1+8\tphi_2-\tphi_3\right)-\frac{f_3}{4}\tphi_3 \, , \nn
\end{align}
whose fundamental matrix is
\beq
\tilde \Upsilon_{\scriptscriptstyle{56}}=\left(\begin{array}{c|c}
 				\frac{1}{r\sqrt{r^4-1}} &\frac{1}{21}\left(-2+3r^4\right)+\frac{2}{21r\sqrt{r^4-1}}\fff(r)\\ \hline
				\frac{1-3r^4}{r^4(r^4-1)} & \frac{2(6r^8-6r^4-1)}{21r^3\sqrt{r^4-1}}+\frac{2(1-3r^4)}{21r^4(r^4-1)}\fff(r)
 			\end{array}\right)\nn\,.
\eeq

A formal solution will have the same structure as~\eqref{formal_phi12}. Recall that $g_{\scriptscriptstyle{56}}^{\phi}$ features quantities defined in terms of one implicit integral coming from $\tphi_1$, $\tphi_2$ and $\tphi_3$. Consequently, the expressions we get are defined in terms of two nested integrals.

The equation for $\tphi_7$ can be cast into the form 
\beq
	\tphi'_7=\tphi'_6-\frac{\ell^6}{m^2}h\,H_0e^{-2u^0}\,\txi_7+\frac{1}{2}f_3\tphi_3-4h^0e^{-2u^0}\tphi_5 \, , \nn
\eeq
where $f_3$ features in equation~\eqref{0th solt}, 
and its solution is given by
\beq
	\tphi_7=\tphi_6-\frac{\ell^6}{m^2}\int h\,H_0e^{-2u^0}\,\txi_7+\frac{1}{2}\int f_3\,\tphi_3-4\int h\,e^{-2u^0}\tphi_5 \, . \nn 
\eeq

Among the summands which appear under integral sign, the first contains no further integral whereas the second integrand is itself defined implicitly and so counts as two nested integrals. The last summand involves three nested integrals (one explicit here and two coming from $\tphi_5$). A simple integration by parts can reduce that number by one, which results in an expression for $\tilde{\phi}_7$ that contains at most two nested integrals. We obtain
\begin{align}
	\tphi_7(r)&=\tphi_6(r)-\frac{\ell^6}{m^2}h\,H_0e^{-2u^0}\,\txi_7(r)+\frac{1}{2}f_3\tphi_3(r)\nonumber\\
		       &\qquad+4\int^r\left(-\frac{2y}{\sqrt{y^4-1}}-2\fff(y)\right)\tphi'_5(y)dy+8\left(\frac{r}{\sqrt{r^4-1}}+\fff(r)\right)\tphi_5(r)\nn\,.
\end{align}

We can now use the $\tphi_1$, $\tphi_2$ system to simplify the equation for $\tilde{\phi}_4$ that is obtained from~\eqref{phi_eqs}, which can be recast to
\begin{align}
	\tphi'_4&=-H^{-1}_0H'_0\,\tphi_4+16\tphi'_1-8\tphi'_2+\frac{1}{2}h\,e^{-2u^0-4v^0}\txi_3-\frac{16m^2}{\ell^6}h\,H^{-1}_0e^{-2u^0-4v^0}f_1\tphi_5\nonumber\\
		   &\qquad -\frac{4m^2}{\ell^6}e^{-4u^0}H^{-1}_0 f_2\tphi_6+\frac{3}{4}\frac{m^2}{\ell^6}e^{-2u^0-4v^0}h\,H^{-1}_0\left(\tphi_6-\tphi_7\right)\label{phi4mas}\,.
\end{align}
The homogeneous solution to this equation is $\tphi_{4,hom}=H^{-1}_0$. Labelling by $g_4^{\phi}$ the non--homogeneous piece of~\eqref{phi4mas}, a general solution is given by
\beq
 	\tilde{\phi}_4(r)=H^{-1}_0(r)Y_4+H^{-1}_0(r)\int^r H_0(y)g_4^{\phi}(y)dy\,.
\eeq

\newpage{}
 \subsubsection{IR asymptotics of the $\phi_a$ modes}
We collect here the IR expansion of the $\phi_a$ fields\footnote{They can are easily obtained from the $\tphi_a$ modes via the inverse transformation~\eqref{inversephi}.}. We write explicitly only the most divergent and constant terms, since higher order terms of the IR expansions (we recall here that the far infrared corresponds to the limit $r\rightarrow 1$ in our conventions) do not provide any constraint on the space of solutions. We also impose throughout the zero energy condition~\eqref{zero_energy_perturb} which requires that $X^{IR}_2=0$.

\begin{align}
	\phi_1&=\frac{1}{\sqrt{r-1}}\Bigg[Y^{IR}_1+\Big(E(-1)-K(-1)\Big)Y_2^{IR}+\frac{\log(r-1)}{4480}\Bigg(-3B_1\Big(34+65K(-1)^2\Big)\nonumber\\
		  &\,\,\,\,\,\,+1792X_1+336X_5-112K(-1)\Big(3X_6-2X_7\Big)\Bigg)\Bigg]+{\cal O}\Big((r-1)^{1/2}\Big)\label{phi1IRsol}\\
	\phi_2&=\frac{1}{13440\sqrt{r-1}}\Bigg[-3B_1\Big(41+100K(-1)^2\Big)+2688X_1+924X_5\nonumber\\
		  &\,\,\,\,\,\,-308K(-1)\Big(3X_6-2X_7\Big)\Bigg]-Y_2^{IR}+{\cal O}\Big((r-1)^{1/2}\Big)\\
	\phi_3&=\frac{1}{15482880 \left(K(-1)^2-4\right)
   \sqrt{r-1}}\Bigg[480  \log (r-1)\left(K(-1)^2-4\right) \Big(3B_1 \left(K(-1)^2+17\right)\nonumber\\&-56 (K(-1) (2 X_7-3X_6)+3 X_5)\Big)-42 K(-1)^2 (21067 B_1-49152
  X_4+17384 X_5)\nonumber\\&+87369 B_1 K(-1)^4-374856 B_1-32 K(-1) \Big(189168 X_6-120117 X_7\nonumber\\&+32 (5210Y_2^{IR}-33 (7 Y_3^{IR}+80 Y_6^{IR}))\Big)+40 K(-1)^3 (36624
   X_6-22535 X_7)\nonumber\\&+1344 \Big(4160 X_1-19200 X_4+311 X_5+160 (7 Y_1^{IR}+7 Y_2^{IR} E(-1)+132 Y_5^{IR}-144 Y_7^{IR})\Big)\Bigg]\nonumber\\
   &+\frac{1}{256} \left(3B_1K(-1)-8X_7\right)\log(r-1)\nonumber\\&-\frac{2 Y_4^{IR}}{3 \left(K(-1)^2-4\right)}+\frac{1}{96}\left(48Y_2^{IR}+Y_3^{IR}\right)+{\cal O}\Big((r-1)^{1/2}\Big)
\end{align}

\begin{align}
	\phi_4&=\frac{1}{7741440
   \left(K(-1)^2-4\right) \sqrt{r-1}}\Bigg[480\log (r-1) \left(K(-1)^2-4\right)  \Big(3B_1 \left(K(-1)^2+17\right)\nonumber\\&-56 (K(-1) (2 X_7-3 X_6)+3 X_5)\Big)+6 K(-1)^2 (9203 B_1-56 (92160
   X_4+6781 X_5))\nonumber\\&-30135 B_1 K(-1)^4-2254920 B_1+32 \Big(K(-1) (488208 X_6-331467 X_7\nonumber\\&+32 (-5210 Y_2^{IR}+231 Y_3^{IR}+2640 Y_6^{IR}))+42\big(4160
   X_1+79104 X_4+4919 X_5\nonumber\\&+160 (7 Y_1^{IR}+7 Y_2^{IR} E(-1)+132 Y_5^{IR}-144 Y_7^{IR})\big)\Big)+8 K(-1)^3 (338909 X_7-494256 X_6)\Bigg]\nonumber\\
   &+\frac{1}{128}\left(3B_1K(-1)-8X_7\right)\log(r-1)-\frac{4 Y_4^{IR}}{3 \left(K(-1)^2-4\right)}+Y_2^{IR}-\frac{5Y_3^{IR}}{16}\nonumber\\&+{\cal O}\Big((r-1)^{1/2}\Big)\\
	\phi_5&=\frac{1}{5160960}\frac{1}{\sqrt{r-1}}\Bigg[60\log (r-1)\left(K(-1)^2-4\right) \Big(3 B_1 \left(K(-1)^2+17\right)\nonumber\\&-56 \big(K(-1) (2 X_7-3 X_6)+3 X_5\big)\Big)+6 K(-1)^2 (1795 B_1-3976
   X_5)+2235 B_1 K(-1)^4\nonumber\\&-42024 B_1-32 K(-1) (6384 X_6-3711 X_7+4160 Y_2^{IR}-672 Y_3^{IR}-7680 Y_6^{IR})\nonumber\\&+152 K(-1)^3 (336 X_6-179
   X_7)+1344 \big(64 X_1-768X_4-23X_5-160 (Y_1^{IR}+Y_2^{IR} E(-1)\nonumber\\&-12 Y_5^{IR})\big)\Bigg]-\frac{3(K(-1)^2-4)}{2048}(3B_1K(-1)-8X_7)+{\cal O}\Big((r-1)^{1/2}\Big)\\
	\phi_6&=\frac{1}{20643840}\Bigg[6 K(-1)^2 (36599B_1+30856 X_5)-5115 B_1 K(-1)^4+140376B_1\nonumber\\
	&-1344\big(1262X_1-2304X_4+185X_5+160(5Y_1^{IR}+12Y_5^{IR}-48Y_7^{IR}+5Y_2^{IR}E(-1))\big)\nonumber\\
	&+32 K(-1) (28560 X_6-18495 X_7+44480 Y_2^{IR}-672Y_3^{IR}-7680 Y_6^{IR})\nonumber\\&+8 K(-1)^3 (16841 
  X_7-26544 X_6)\Bigg]+{\cal O}\Big((r-1)\Big)
\end{align}

\begin{align}
	\phi_7&=\frac{1}{5160960(r-1)}\Bigg[6K(-1)^2(5656X_5-295B_1)-8K(-1)^3(7644X_6-4241X_7)\nonumber\\
		  &-2415B_1K(-1)^4+32K(-1)\big(7644X_6-4551X_7+32(130Y_2^{IR}-21Y_3^{IR}-240Y_6^{IR})\big)\nonumber\\
		  &+8904B_1-1344\big(64X_1-768X_4+7X_5-160(Y_1^{IR}-12Y_5^{IR}+Y_2^{IR}E(-1))\big)\Bigg]\nonumber\\
		  &-\frac{\log(r-1)}{(r-1)}\,\frac{K(-1)^2-4}{86016}\Bigg[3B_1(17+K(-1)^2)-56\big(3X_5-K(-1)(3X_6-2X_7)\big)\Bigg]\nonumber\\
		  &+\frac{\log(r-1)}{860160}\Bigg[B_1(3468+4485K(-1)^2-15K(-1)^4)-56\big(768X_1+204X_5\nonumber\\
		  &-68K(-1)(3X_6-2X_7)-15K(-1)^2X_5+5K(-1)^3(3X_6-2X_7)\big)\Bigg]\nonumber\\
		  &+\frac{1}{20643840}\Bigg[32K(-1)\big(28650X_6-18495X_7+32(1390Y_2^{IR}-21Y_3^{IR}-240Y_6^{IR})\big)\nonumber\\
		  &+6K(-1)^2(36599B_1+30856X_5)-8K(-1)^3(26544X_6-16841X_7)-5115K(-1)^4B_1\nonumber\\
		  &+140376B_1-1344\big(1216X_1-2304X_4+181X_5\nonumber\\
		  &+160(5Y_1^{IR}+12Y_5^{IR}+48Y_7+5Y_2^{IR}E(-1))\big)\Bigg]+{\cal O}\Big((r-1)^{1/2}\Big)\label{phi7IRsol}
\end{align}

\section{Elliptic functions}\label{ellfunc}

For the reader's convenience, we list in this section the definitions of elliptic functions of which we make frequent use in the bulk of the text and especially in Appendix A.
The incomplete elliptic integral of the first kind is defined as
\begin{align}
F(\phi,m)&=\int^{\phi}_{0}\frac{\mbox{d}\theta}{\sqrt{1- m^2\, \mbox{sin}^2\, \theta}}\,\,
\end{align}
while the complete elliptic integral of the first kind reads
\beq
K(m)=F\,(\frac{\pi}{2},m)\,.
\eeq
Analogously, $E(\phi,m)$, the incomplete elliptic integral of the second kind, has the following expression
\beq
E(\phi,m)=\int^{\phi}_0\sqrt{1-m^2\, \mbox{sin}^2\, \theta}\, d\theta\,,
\eeq
and is related to the complete elliptic integral of the second kind as
\beq
E(m)=E\,(\frac{\pi}{2},m)\,.
\eeq

\end{appendix}



\end{document}